\documentclass[a4paper,11pt,reqno]{amsart}

\usepackage{amssymb,amsmath}
\usepackage{mathrsfs}

\newtheorem{thm}{Theorem}[section]
\newtheorem{prp}[thm]{Proposition}
\newtheorem{lem}[thm]{Lemma}
\newtheorem{dfn}[thm]{Definition}
\newenvironment{defn}{\begin{dfn} \rm }{\end{dfn}}
\newtheorem{cor}[thm]{Corollary}
\newtheorem{example}[thm]{Example}
\newenvironment{exa}{\begin{example} \rm }{ \end{example}}
\newtheorem{remark}[thm]{Remark}

\newenvironment{rmk}{\begin{remark} \rm }{\hfill $\Box$ \end{remark}}
\newenvironment{prf}{\noindent {\it Proof:} \ }{\hfill $\Box$}
\newenvironment{prfof}[1]{\noindent {\it Proof of #1} \ }{\hfill $\Box$}

\newcommand\od{\mathrm{d}}
\newcommand\p{\partial}

\newcommand{\nn}{\nonumber}

\newcommand{\ld}{\lambda}

\newcommand{\dt}{\delta}
\newcommand{\ta}{\theta}

\newcommand{\ve}{\varepsilon}

\newcommand{\res}{\mathrm{Res}}

\newcommand\Z{\mathbb{Z}}
\newcommand\C{\mathbb{C}}

 \newcommand\cB{\mathcal{B}}

\newcommand\cE{\mathcal{E}}

\newcommand{\set}[1]{\left\{#1\right\}}

\newcommand\ra{\right\rangle}
\newcommand\la{\left\langle}

\newcommand\Zop{\mathbb{Z}^{\mathrm{odd}}_{>0} }
\newcommand{\bm}[1]{\mathbf{#1}}
\newcommand{\bs}[1]{\boldsymbol{#1}}
\newcommand{\bt}{\bm{t}}  

\allowdisplaybreaks \numberwithin{equation}{section}

\begin{document}

\title[Extension of the KP hierarchy via 
operators with two derivations]{A KP-mKP Hierarchy via Operators with Two Derivations}
\author{Lumin Geng, Jianxun Hu, Chao-Zhong Wu}
\dedicatory {School of Mathematics, Sun Yat-sen University, Guangzhou 510275, P. R. China \\
Email address: genglm@mail2.sysu.edu.cn, stsjxhu@sysu.edu.cn,  wuchaozhong@sysu.edu.cn}

\begin{abstract}
By using pseudo-differential operators containing two derivations, we extend the Kadomtsev-Petviashvili (KP) hierarchy to a certain KP-mKP hierarchy. For the KP-mKP hierarchy, we obtain its B\"{a}cklund transformations, bilinear equations of Baker-Akhiezer functions and Hirota equations of tau functions. Moreover, we show that this hierarchy is equivalent to a subhierarchy of the dispersive Whitham hierarchy associated to the Riemann sphere with its infinity point and one movable point marked.
\end{abstract}
\maketitle

\section{introduction}

The Kadomtsev-Petviashvili (KP) hierarchy and the modified KP (mKP) hierarchy \cite{JM1983}, as two fundamental models in the theory of integrable systems, have been investigated from various points of view. For instance, these two integrable hierarchies can be converted to each other via certain gauge transformations (see \cite{OWRC1993} and references therein), and their tau functions have important application in matrix models as well as intersection numbers on the moduli space of Riemann surfaces (see, for example, \cite{Ale,Bur2015,Bur2016,Kon,Wit}).

In the present paper, we are to construct an integrable hierarchy that can be reduced to either the KP hierarchy or to the mKP hierarchy, with the help of pseudo-differential operators of two derivations. More exactly, let us consider two pseudo-differential operators of the form
\begin{align*}
\Phi_1=1+\sum_{i\ge1}a_{1,i} \p_{1}^{-i},\quad
\Phi_2=e^{\beta }\Big(1+\sum_{i\ge1}a_{2,i}\p_{2}^{-i}\Big),
\end{align*}
where $a_{\nu,i}$ and $\beta$ are smooth functions of the variables $x$ and $y$, and $\p_1=\p/\p x$ and $\p_2=\p/\p y$ are derivations commuting with each other. We introduce the following evolutionary equations:
\begin{align}
	\frac{\p \Phi_1}{\p t_{1, k}}=-\left(\Phi_1\p_1^k\Phi_1^{-1}\right)_{<0}\Phi_1,&\quad
	\frac{\p \Phi_2}{\p t_{2, k}}=-\left(\Phi_2\p_2^k\Phi_2^{-1}\right)_{<1}\Phi_2, \label{Phinut10} \\
	\frac{\p \Phi_1}{\p t_{2, k}}=\left(\Phi_2\p_2^k\Phi_2^{-1}\right)_{\geq1}\la\Phi_1\ra,&\quad
	\frac{\p \Phi_2}{\p t_{1, k}}=\left(\Phi_1\p_1^k\Phi_1^{-1}\right)_{\geq0}\la\Phi_2\ra, \label{Phinut20} \\
e^\beta\p_1 e^{-\beta}\Phi_1\p_2\Phi_1^{-1}&=\p_2\Phi_2\p_1\Phi_{2}^{-1} \label{Phinumu0}
\end{align}
with  $k=1,2,3,\cdots$.
Here it is employed the notations for truncations of pseudo-differential operators as:
\[
\left(\sum_{i}f_i\p_\nu^i\right)_{i\ge m}=\sum_{i\ge m}f_i\p_\nu^i, \quad
\left(\sum_{i}f_i\p_\nu^i\right)_{i< m}=\sum_{i<m}f_i\p_\nu^i,
\]
and for actions of a differential operator $D$ as
\[
D\la\sum_{i}f_i\p_\nu^i \ra=\sum_{i} D(f_i)\p_\nu^i.
\]
The equations \eqref{Phinut10}--\eqref{Phinumu0} will be proved compatible in Section 3, hence they indeed compose an integrable hierarchy. In particular, the equations $\p\Phi_1/\p t_{1,k}$ form the KP hierarchy, while the equations $\p\Phi_2/\p t_{2,k}$ form the mKP hierarchy. For this reason, we just call the system of equations \eqref{Phinut10}--\eqref{Phinumu0} the \emph{KP-mKP hierarchy}.

In the KP-mKP hierarchy \eqref{Phinut10}--\eqref{Phinumu0}, one observes that
\[
\frac{\p e^\beta}{\p t_{1,k} }=\left(\Phi_1\p_1^k\Phi_1^{-1}\right)_{\geq0}(e^\beta).
\]
Namely, the function $e^\beta$
plays the role of eigenfunction of the KP hierarchy. Such an eigenfunction induces a
B\"{a}cklund transformation of the KP-mKP hierarchy (see Proposition~\ref{thm-invol} below)
\begin{equation*}\label{BT0}
\iota~:~(\Phi_1,\Phi_2)\mapsto\left.\left(e^{-\beta}\Phi_2,e^{-\beta}\Phi_1\right) \right|_{(\bt_1,\bt_2)\mapsto(\bt_2,\bt_1);\, (\p_1,\p_2)\mapsto(\p_2,\p_1)},
\end{equation*}
where $\bt_\nu=(t_{\nu,1},t_{\nu,2},t_{\nu,3},\dots)$. From these B\"{a}cklund transformations one recovers the gauge transformations between the KP and the mKP hierarchies studied in \cite{OWRC1993}.
It can also be seen that, if we take the eigenfunction $e^\beta=1$, and assume $\Phi_{\nu}^*=\p_\nu\Phi_\nu^{-1}\p_\nu^{-1}$ for $\nu\in\{1,2\}$, then the KP-mKP hierarchy with $k\in\Z^{\mathrm{odd}}_+$ is reduced to the two-component BKP hierarchy, which is deeply related to representation theory and algebraic geometry \cite{CM,DJKM-KPtype,GHW2023,LWZ2011,Shi}.

Given a solution of the KP-mKP hierarchy \eqref{Phinut10}--\eqref{Phinumu0}, we introduce its Baker-Akhiezer functions $w_\nu(\bt_1,\bt_2;z)$ and its adjoint Baker-Akhiezer functions $w_\nu^\dag(\bt_1,\bt_2;z)$, where $\nu\in\{1,2\}$ and $z$ is a parameter. Then the KP-mKP hierarchy will be seen equivalent to the following bilinear equation (see Theorem~\ref{thm-ble2} below)
\begin{equation}\label{ble0}
	\res_z\left( z^{-1} w_1(\bt_1, \bt_2; z)w_1^{\dag}(\bt_1', \bt_2'; z) \right)=\res_z \left(z^{-1}w_2(\bt_1, \bt_2; z)w_2^{\dag}(\bt_1', \bt_2'; z)\right)
\end{equation}
for arbitrary time variables $(\bt_1, \bt_2)$ and $(\bt'_1, \bt'_2)$. With a method similar as that in \cite{DKV2022,Dickey,GHW2023}, we will define two tau functions $\tau_1(\bt_1, \bt_2)$ and $\tau_2(\bt_1, \bt_2)$ of the KP-mKP hierarchy. Furthermore, we will see that these tau functions are related to the (adjoint) Baker-Akhiezer functions via the formulae:
\begin{align*} w_1(\bt_1,\bt_2;z)&=\frac{\tau_1(\bt_1-[z^{-1}],\bt_2)}{\tau_1(\bt_1,\bt_2)}e^{\xi(\bt_1;z)},\quad w_1^{\dag}(\bt_1,\bt_2;z)=\frac{\tau_2(\bt_1+[z^{-1}],\bt_2)}{\tau_2(\bt_1,\bt_2)}e^{-\xi(\bt_1;z)},\\ w_2(\bt_1,\bt_2;z)&=\frac{\tau_2(\bt_1,\bt_2-[z^{-1}])}{\tau_1(\bt_1,\bt_2)}e^{\xi(\bt_2;z)},\quad w_2^{\dag}(\bt_1,\bt_2;z)=\frac{\tau_1(\bt_1,\bt_2+[z^{-1}])} {\tau_2(\bt_1,\bt_2)}e^{-\xi(\bt_2;z)}, 
\end{align*}
where $\xi(\bt_\nu;z)=\sum_{k\ge1}t_{\nu,k}z^k$, and $[z^{-1}]=\left(\frac{1}{z},\frac{1}{2 z^2},\frac{1}{3 z^3},\dots\right)$. With the help of these formulae, the bilinear equation \eqref{ble0} will be recast to the form of Hirota equations \cite{Hir} of $\tau_1$ and $\tau_2$. In particular, we can write down the Hirota equations for the mKP hierarchy \cite{JM1983}, which were conjectured by Alexandrov \cite{Ale} to govern the generating function of the intersection
numbers on the moduli spaces of Riemann surfaces with boundary (see also \cite{Bur2015,Bur2016}).

Our motivation is also from the study of the Whitham hierarchy. In 1994, Krichever introduced the universal Whitham hierarchy on the moduli spaces of Riemann surfaces of arbitrary genus with $N$ marked points, and revealed its relation with the Witten-Dijgraagh-Verlinder-Verlinder (WDVV) equations \cite{KIM1994}. In the case of genus zero, a dispersive version of the Whitham hierarchy was proposed by Szablikowski and Blaszak \cite{SB2008} in the form of Lax equations of operators containing a single derivation. In particular, the Lax equations corresponding to the infinity point $\infty$ give the KP hierarchy, and those corresponding to $\infty$ together with a movable point give an extension of the KP hierarchy. Such an extended KP hierarchy was further studied by Lu, Wu and Zhou \cite{LW2021,WZ2016}, who obtained its bi-Hamiltonian structures, bilinear equation of (adjoint) Baker-Akhiezer functions and additional symmetries. Moreover, underlying this hierarchy there exists a class of infinite-dimensional Dubrovin-Frobenius manifolds \cite{MWZ2021}. In the present paper, we will represent the KP-mKP hierarchy under certain generic condition into another version of bilinear equation (see \eqref{exble} below) in contrast to the one \eqref{ble0}, and show that it coincides with the bilinear equation of the extended KP hierarchy given in \cite{LW2021}. In other words, the KP-mKP hierarchy is proved equivalent to a subhierarchy of the dispersive Whitham hierarchy associated to the Riemann sphere with $\infty$ and one movable point marked. In a sense, the B\"{a}cklund transformation \eqref{BT0} is illustrated by exchanging $\infty$ and the movable point on the Riemann sphere.

This paper is organized as follows. In the next section, we layout some notations of pseudo-differential operators containing two derivations. With the help of these operators, we prove that the KP-mKP hierarchy \eqref{Phinut10}--\eqref{Phinumu0} is well defined in Section~3, and derive its B\"{a}cklund transformations as well as some other properties in Section~4. In Section~5, we rewrite the KP-mKP hierarchy into two versions of bilinear equations of (adjoint) Baker-Akhiezher functions, one of which is equivalent to the extended KP hierarchy studied in \cite{SB2008,WZ2016}. In Section~6, we introduce two tau functions of the KP-mKP hierarchy, and obtain a system of Hirota equations satisfied by them. The final section is devoted to some remarks.

\section{Pseudo-differential operators with two derivations}

Let us recall the notion of pseudo-differential operators containing two derivations, following \cite{GHW2023}.
Let $\cB$ be a commutative associative algebra of smooth complex functions of two variables $x$ and $y$.  The derivations $\p_1=\od/\od x$ and $\p_2=\od/\od y$ act on $\cB$, and they are assumed to commute with each other. Let us consider a set as follows
\begin{align*}\label{}
&\cE=\left\{ \sum_{i\le m}\sum_{j\le n}f_{i j} \p_1^i \p_2^j\mid f_{i j}\in\cB; \, m,n\in\Z \right\}.
\end{align*}
An element of $\cE$ is called a pseudo-differential operator with two derivations $\p_1$ and $\p_2$.
One sees that $\cE$ is an associative algebra over $\C$ with a product defined by 
\begin{equation*}\label{}
f \p_1^i\p_2^j\cdot g \p_1^p \p_2^q=\sum_{r,s\geq0}\binom{i}{r}\binom{j}{s} f\, \p_1^r\p_2^s(g)\cdot
\p_1^{i+p-r}\p_2^{j+q-s},
\end{equation*}
where $f,\,g\in\cB$ and
\[
\binom{i}{r}=\frac{i(i-1)\dots(i-r+1)}{r!}.
\]
In $\cE$ a Lie bracket is defined by the commutator, that is,
\[
[A, B]=A B-B A, \quad A, B\in\cE.
\]

Let $A=\sum_{i\le m}\sum_{j\le n}  {f}_{i j} \p_1^i\p_2^j$ be an arbitrary operator in $\cE$. We have the following projections:
\begin{equation}\label{proj}
A_+=\sum_{0\le i\le m}\sum_{0\le j\le n} f_{i j} \p_1^i\p_2^j, \quad A_-=\sum_{i\le 0;\, j\le 0; \, i+j<0 } f_{i j} \p_1^i\p_2^j.
\end{equation}
The residues of $A$ mean
\begin{equation*}\label{resA}
\res_{\p_1}A=\sum_{j\le n}  {f}_{-1, j} \p_2^j, \quad \res_{\p_2}A=\sum_{i\le m}  {f}_{i, -1} \p_1^i,
\end{equation*}
and its adjoint operator is
\begin{equation*}\label{star}
A^*=\sum_{i\le m}\sum_{j\le n} (-1)^{i+j}  \p_1^i\p_2^j f_{i j}.
\end{equation*}
It is easy to see
\begin{equation*}
(\res_{\p_\nu} A)^*=-\res_{\p_\nu} A^*, \quad \nu\in\set{1,2}.
\end{equation*}
Moreover, $``*"$ is an anti-automorphism on $\mathcal{E}$, namely, for any $B, C\in\cE$,
\[
(B^*)^*=B, \quad (B C)^*=C^* B^*.
\]
Given a differential operator $D=\sum_{r,s\ge0}  {g}_{r s} \p_1^r\p_2^s\in\cE$, its action on $A$ is denoted by
\begin{equation*}\label{exDA}
D\la{A}\ra=\sum_{i,j} \left(\sum_{r,s\ge0}  {g}_{r s} \p_1^r \p_2^s (f_{i j})\right) \p_1^i\p_2^j,
\end{equation*}
that is, the differential operator acts on each coefficient of $A$.
Here note that $g_{r s}\ne0$ for only finitely many $r$ and $s$.

Clearly, whenever the operator $A$ contains powers in only $\p_1$ or $\p_2$, then the above notations agree with those for pseudo-differential operators involving a single derivation. In this case, for $A=\sum_{i}f_i\p_\nu^i\in\mathcal{E}$ and $m\in\Z$, we will use the following truncations:
\begin{equation*}\label{cutop}
A_{\ge m}=\sum_{i\ge m} f_i \p_\nu^i, \quad A_{< m}=\sum_{i<m} f_i \p_\nu^i.
\end{equation*}
For instance, in this case $A_{\ge0}=A_+$ and $A_{<0}=A_-$ in consideration of \eqref{proj}.

\section{Lax equations via pseudo-differential operators with two derivations}

In this section we want to introduce an integrable hierarchy via pseudo-differential operators with two derivations.

Let us start with two pseudo-differential operators in $\cE$ of the form
\begin{align}
\Phi_1=1+\sum_{i\ge1}a_{1,i} \p_{1}^{-i},\quad
\Phi_2=e^{\beta }\Big(1+\sum_{i\ge1}a_{2,i}\p_{2}^{-i}\Big),\label{exPhi12}
\end{align}
where $a_{\nu,i}$, $\beta\in\cB$. 
It is easy to see that these two operators are invertible, hence the following  pseudo-differential operators make sense
\begin{equation*}\label{exPnu}
P_\nu=\Phi_\nu\p_\nu\Phi_\nu^{-1}, \quad \nu\in\set{1,2}.
\end{equation*}
More precisely, these two operators take the form
\begin{align}
	P_1=\p_1+\sum_{i\ge1}u_{1,i}\p_{1}^{-i},\quad
	P_2=\p_2-\p_2(\beta)+\sum_{i\ge1}u_{2,i}\p_{2}^{-i}, \label{exP2}
\end{align}
where $u_{\nu,i}$ are differential polynomials belonging to the ring
\[
\mathcal{R}_\nu=\C\left[ \dt_{\nu 2}\p_\nu^m(\beta), \p_\nu^m(a_{\nu,k}) \mid m\ge0; k\ge1 \right].
\]
Here and below $\dt$ stands for the Kronecker delta symbol. In particular, one has
\begin{equation}\label{u12}
u_{\nu,1}=-\p_\nu\left(a_{\nu,1}\right), \quad \nu\in\{1,2\}.
\end{equation}

Denote $\bt_\nu=\left\{t_{\nu,1}, t_{\nu,2}, t_{\nu,3}, \dots\right\}$ for $\nu\in\{1,2\}$. Assume the coefficients $a_{\nu,k}$ and $\beta$ in \eqref{exPhi12} to be unknown functions depending on the variable $(\bt_1, \bt_2)$, and they satisfy the following evolutionary equations:
\begin{align}
\frac{\p \Phi_1}{\p t_{1, k}}=-(P_1^k)_{<0}\Phi_1,&\quad
\frac{\p \Phi_2}{\p t_{2, k}}=-(P_2^k)_{<1}\Phi_2, \label{exPhinut1} \\
\frac{\p \Phi_1}{\p t_{2, k}}=(P_2^k)_{\geq1}\la\Phi_1\ra,& \quad
 \frac{\p \Phi_2}{\p t_{1, k}}=(P_1^k)_{\geq0}\la\Phi_2\ra, \label{exPhinut2} \\
e^\beta\p_1 e^{-\beta}\Phi_1\p_2\Phi_1^{-1}&=\p_2\Phi_2\p_1\Phi_{2}^{-1}, \label{exPhinumu}
\end{align}
where
$k\in\Z_{\geq1}$. What is more, these derivations $\p/\p t_{\nu,k}$ are assumed to commute with $\p_1$ and $\p_2$. In particular, since ${\p}/{\p t_{\nu,1}}=\p_{\nu}$, then in what follows we will just take
\[
t_{1,1}=x, \quad t_{2,1}=y.
\]

\begin{thm}\label{thm-welldef}
The evolutionary equations \eqref{exPhinut1}--\eqref{exPhinumu} are well defined, and they are compatible.
\end{thm}

Let us proceed to prove the theorem. To this end, we need to show that the flows \eqref{exPhinut1}--\eqref{exPhinut2} commute with each other, and that the constraint \eqref{exPhinumu} is invariant with respect to such flows.

Firstly, we introduce the following notations:
\begin{equation}\label{exBk}
B_{\nu,k}^\mu=\left\{ \begin{array}{cl}
                        -(P_\nu^k)_{<\nu-1}, & \mu=\nu; \\
                        (P_\nu^k)_{\geq\nu-1}\la\Phi_\mu\ra \Phi_\mu^{-1}, & \mu\ne\nu.
                      \end{array}\right.
\end{equation}
Here $\mu,\nu\in\set{1,2}$ and $k\in\Z_{\geq1}$, and these indices mean the same below unless otherwise stated.
It is easy to see that the operators $B_{\nu,k}^\mu$ are pseudo-differential operators with a single derivation $\p_\mu$, and that $(B_{\nu,k}^\mu)_{\geq\mu-1}=0$. With the help of these notions, the equations \eqref{exPhinut1}--\eqref{exPhinut2} can be rewritten as
\begin{equation}\label{exPhiBk}
\frac{\p\Phi_\mu}{\p t_{\nu,k}}=B_{\nu,k}^\mu\Phi_\mu,
\end{equation}
which implies
\begin{equation}\label{exPnut}
\frac{\p P_\mu}{\p t_{\nu,k}}=\left[B_{\nu,k}^\mu, P_\mu\right]. 
\end{equation}

\begin{lem}\label{thm-exBkt}
The operators \eqref{exBk} satisfy the following equalities
\begin{equation*}\label{exBkt}
\frac{\p
B_{\nu,l}^\ld}{\p t_{\mu,k} }-\frac{\p
B_{\mu,k}^\ld}{\p t_{\nu,l} }+\left[B_{\nu,l}^\ld, B_{\mu,k}^\ld \right]=0,
\end{equation*}
where $\ld, \mu, \nu\in\set{1,2}$, $k, l\in\Z_{\geq1}$.
\end{lem}
\begin{prf}
For convenience, the left hand side of \eqref{exBkt} is denoted as ``l.h.s.''. Let us calculate it case by case with the help of \eqref{exBk}--\eqref{exPnut}.
\begin{itemize}
\item[(i)] When $\mu=\nu=\ld$, we have
\begin{align*}
  \hbox{l.h.s.}=& \frac{\p }{\p t_{\nu,k} }(-P_\nu^l)_{<\nu-1} -\frac{\p}{\p t_{\nu,l} }(-P_\nu^k)_{<\nu-1} +\left[-(P_\nu^l)_{<\nu-1}, -(P_\nu^k)_{<\nu-1} \right]
  \\
  =&  \left[-(P_\nu^k)_{<\nu-1}, -P_\nu^l  \right]_{<\nu-1} - \left[-(P_\nu^l)_{<\nu-1}, -P_\nu^k  \right]_{<\nu-1} \\
  &+\left[-(P_\nu^l)_{<\nu-1}, -(P_\nu^k)_{<\nu-1}\right]_{<\nu-1}
  \\
  =&  \left[-(P_\nu^k)_{<\nu-1}, -P_\nu^l  \right]_{<\nu-1} - \left[-(P_\nu^l)_{<\nu-1}, -(P_\nu^k)_{\ge\nu-1}  \right]_{<\nu-1} \\
   =&  \left[-(P_\nu^k)_{<\nu-1}, -P_\nu^l  \right]_{<\nu-1} + \left[-(P_\nu^k)_{\ge\nu-1}, -P_\nu^l   \right]_{<\nu-1} \\
    =&\left[-P_\nu^k,-P_\nu^l\right]_{<\nu-1}=0.
\end{align*}
\item[(ii)] When $\mu=\nu\ne\ld$, we have
\begin{align*}
  \hbox{l.h.s.}=& \frac{\p }{\p t_{\nu,k} }\left( (P_\nu^l)_{\geq\nu-1}\la\Phi_\ld\ra\Phi_\ld^{-1}\right) -(k\leftrightarrow l) \\
   &+\left[ (P_\nu^l)_{\geq\nu-1}\la\Phi_\ld\ra\Phi_\ld^{-1},  (P_\nu^k)_{\geq\nu-1}\la\Phi_\ld\ra\Phi_\ld^{-1} \right]  \\
  =&  \left[-(P_\nu^k)_{<\nu-1}, P_\nu^l\right]_{\geq\nu-1}\la\Phi_\ld\ra\Phi_\ld^{-1}
  + (P_\nu^l )_{\geq\nu-1}\la(P_\nu^k )_{\geq\nu-1}\la\Phi_\ld\ra\ra\Phi_\ld^{-1}-(k\leftrightarrow l)\\
  =&\left(\left[-(P_\nu^k)_{<\nu-1},P_\nu^l\right]_{\geq\nu-1} -\left[-(P_\nu^l)_{<\nu-1},P_\nu^k\right]_{\geq\nu-1} \right. \\ &\left.+\left[(P_\nu^l)_{\geq\nu-1},(P_\nu^k)_{\geq\nu-1}\right]_{\geq\nu-1}\right)\la\Phi_\ld\ra\Phi_\ld^{-1}\\
  =&\left(\left[P_\nu^l,(P_\nu^k)_{<\nu-1}\right]_{\geq\nu-1} +\left[P_\nu^l,(P_\nu^k)_{\geq\nu-1}\right]_{\geq\nu-1}\right)\la\Phi_\ld\ra\Phi_\ld^{-1} \\
  =& \left[P_\nu^l,P_\nu^k\right]_{\geq\nu-1}\la\Phi_\ld\ra\Phi_\ld^{-1}=0.
\end{align*}
Here the notation ``$(k\leftrightarrow l)$'' stands for all the terms to the left of it with the indices $k$ and $l$ exchanged.
\item[(iii)]  When $\mu\ne\nu=\ld$, we have
\begin{align}
  \hbox{l.h.s.}=& \frac{\p }{\p t_{\mu,k} }(-P_\nu^l)_{<\nu-1} -\frac{\p}{\p t_{\nu,l} }\left( (P_\mu^k)_{\geq\mu-1}\la\Phi_\nu\ra\Phi_\nu^{-1}\right) \nn\\
   &+\left[-(P_\nu^l)_{<\nu-1}, (P_\mu^k)_{\geq\mu-1}\la\Phi_\nu\ra\Phi_\nu^{-1} \right]
  \nn\\
  =& \left[(P_\mu^k)_{\geq\mu-1}\la\Phi_\nu\ra\Phi_\nu^{-1}, -P_\nu^l \right]_{<\nu-1}  -\left[(P_\nu^l)_{\geq\nu-1}\la\Phi_\mu\ra\Phi_\mu^{-1}, P_\mu^k \right]_{\geq\mu-1} \la\Phi_\nu\ra\Phi_\nu^{-1} \nn\\
  &   - (P_\mu^k)_{\geq\mu-1} \la -(P_\nu^l)_{<\nu-1}\Phi_\nu\ra\Phi_\nu^{-1}
  + (P_\mu^k)_{\geq\mu-1}\la \Phi_\nu\ra\Phi_\nu^{-1}(-(P_\nu^l)_{<\nu-1}) \nn\\  &+\left[(P_\mu^k)_{\geq\mu-1}\la\Phi_\nu\ra\Phi_\nu^{-1}, (P_\nu^l)_{<\nu-1} \right]_{<\nu-1} \nn\\
  =& -\left[\res_{\p_{\mu}}(P_\mu^k)_{\geq\mu-1}\Phi_\nu\p_{\mu}^{-1}\Phi_\nu^{-1}, (P_\nu^l)_{\geq\nu-1} \right]_{<\nu-1}  \nn\\  & -\left[(P_\nu^l)_{\geq\nu-1}\la\Phi_\mu\ra\Phi_\mu^{-1}, P_\mu^k \right]_{\geq\mu-1} \la\Phi_\nu\ra\Phi_\nu^{-1} \nn\\
  & +\res_{\p_\mu}\left( (P_\mu^k)_{\geq\mu-1} (P_\nu^l)_{<\nu-1}\Phi_\nu\p_{\mu}^{-1}\Phi_\nu^{-1}   - (P_\mu^k)_{\geq\mu-1} \Phi_\nu\p_{\mu}^{-1}\Phi_\nu^{-1}(P_\nu^l)_{<\nu-1} \right) \nn\\
   =
   &-\left(\res_{\p_{\mu}}\left[(P_\mu^k)_{\geq\mu-1}, (P_\nu^l)_{\geq\nu-1} \right]\Phi_\nu\p_{\mu}^{-1}\Phi_\nu^{-1}\right)_{<\nu-1} \nn\\
   &-\left(\res_{\p_{\mu}}(P_\mu^k)_{\geq\mu-1}\left[\Phi_\nu\p_{\mu}^{-1}\Phi_\nu^{-1}, (P_\nu^l)_{\geq\nu-1} \right]\right)_{<\nu-1} \nn\\
   &  -\left[(P_\nu^l)_{\geq\nu-1}\la\Phi_\mu\ra\Phi_\mu^{-1}, P_\mu^k \right]_{\geq\mu-1} \la\Phi_\nu\ra\Phi_\nu^{-1}  \nn\\
   &+\res_{\p_\mu}\left( (P_\mu^k)_{\geq\mu-1}\left[ (P_\nu^l)_{<\nu-1},\Phi_\nu\p_{\mu}^{-1}\Phi_\nu^{-1}  \right] \right)_{<\nu-1} \nn\\
=
&-\left(\res_{\p_{\mu}}\left[(P_\mu^k)_{\geq\mu-1}, (P_\nu^l)_{\geq\nu-1} \right]\Phi_\nu\p_{\mu}^{-1}\Phi_\nu^{-1}\right)_{<\nu-1}  \nn\\
 &+\res_{\p_\mu}\left( (P_\mu^k)_{\geq\mu-1}\left[ P_\nu^l,\Phi_\nu\p_{\mu}^{-1}\Phi_\nu^{-1}  \right] \right)_{<\nu-1}  \nn\\
 &-\left[(P_\nu^l)_{\geq\nu-1}\la\Phi_\mu\ra\Phi_\mu^{-1}, P_\mu^k \right]_{\geq\mu-1} \la\Phi_\nu\ra\Phi_\nu^{-1}\nn\\
 =
 &-\left(\res_{\p_{\mu}}\left[(P_\mu^k)_{\geq\mu-1}, (P_\nu^l)_{\geq\nu-1} \right]\Phi_\nu\p_{\mu}^{-1}\Phi_\nu^{-1}\right)_{<\nu-1} \nn\\
 &-\left[(P_\nu^l)_{\geq\nu-1}\la\Phi_\mu\ra\Phi_\mu^{-1}, P_\mu^k \right]_{\geq\mu-1} \la\Phi_\nu\ra\Phi_\nu^{-1}
 . \label{exlhs3}
\end{align}
Here the last equality holds for
\[
 [P_\nu^l, \Phi_\nu\p_{\mu}^{-1}\Phi_\nu^{-1} ] = \Phi_\nu [\p_\nu^l, \p_{\mu}^{-1}]\Phi_\nu^{-1}=0.
\]
It is easy to see that \eqref{exlhs3} is a pseudo-differential operator of a single derivation $\p_\nu$, and that
\[
(\hbox{l.h.s.})_{\geq\nu-1}= -\left(\left[(P_\nu^l)_{\geq\nu-1}\la\Phi_\mu\ra\Phi_\mu^{-1}, P_\mu^k \right]_{\geq\mu-1} \la\Phi_\nu\ra\Phi_\nu^{-1}\right)_{\geq\nu-1}=0.
\]
Moreover, by using the fact
\[
\p_{\nu}^m\la\Phi_\mu\ra=\p_{\nu}^m \Phi_\mu -\sum_{r=1}^{m} \binom{m}{r}\p_{\nu}^{m-r}\la\Phi_\mu\ra\p_{\nu}^r, \quad m\ge1,
\]
one has
\begin{equation*}\label{exPkPhi}
(P_\nu^l)_{\geq\nu-1}\la\Phi_\mu\ra=(P_\nu^l)_{\geq\nu-1}\Phi_\mu + R\p_\nu
\end{equation*}
for a certain operator $R$ of the form
\[R=\sum_{i=0}^{n}\sum_{j\leq0}f_{i j}\p_\nu^i\p_\mu^j,\quad f_{i j}\in\cB.
\]
Accordingly, we have
\begin{align}
&\left( \Big[(P_\nu^l)_{\geq\nu-1}\la\Phi_\mu\ra\Phi_\mu^{-1}, P_\mu^k \right]_{\geq\mu-1}\la\Phi_\nu\ra\Phi_\nu^{-1}\Big)_{<\nu-1}
\nn\\
=&\left(\res_{\p_\mu} \Big(\left[(P_\nu^l)_{\geq\nu-1}\la\Phi_\mu\ra\Phi_\mu^{-1}, P_\mu^k \right]\Phi_\nu\p_{\mu}^{-1}\Phi_\nu^{-1}\Big) \right.\nn\\
& \left.-\dt_{\mu,2}\res_{\p_\mu}\Big(\left[(P_\nu^l)_{\geq\nu-1}\la\Phi_\mu\ra\Phi_\mu^{-1}, P_\mu^k\right]\p_\mu^{-1}\Big) \right)_{<\nu-1}
\nn\\
=&\left(\res_{\p_\mu} \left[(P_\nu^l)_{\geq\nu-1} + R\p_\nu\Phi_\mu^{-1}, P_\mu^k \right]\Phi_\nu\p_{\mu}^{-1}\Phi_\nu^{-1}\right)_{<\nu-1}
\nn\\
=&\left(\res_{\p_\mu} \left[ (P_\nu^l)_{\geq\nu-1}, P_\mu^k \right]\Phi_\nu\p_{\mu}^{-1}\Phi_\nu^{-1}\right)_{<\nu-1}
\nn\\
&+ \left(\res_{\p_\mu}  \left[R\p_\nu\Phi_\mu^{-1}, P_\mu^k \right]  \Phi_\nu\p_{\mu}^{-1}\Phi_\nu^{-1}\right)_{<\nu-1} \nn\\
=&\left(\res_{\p_\mu} \left[ (P_\nu^l)_{\geq\nu-1}, (P_\mu^k)_{\geq\mu-1} \right]\Phi_\nu\p_{\mu}^{-1}\Phi_\nu^{-1} \right)_{<\nu-1} \nn\\
&+\left(\left[ (P_\nu^l)_{\geq\nu-1}, \dt_{\mu,2}\res_{\p_\mu}(P_\mu^k\p_\mu^{-1}) \right]
\right)_{<\nu-1} \nn\\
&+\left(\res_{\p_\mu} \left( R\p_\nu\Phi_\mu^{-1} P_\mu^k  -P_\mu^k R\p_\nu \Phi_\mu^{-1} \right) \Phi_\nu\p_{\mu}^{-1}\Phi_\nu^{-1}\right)_{<\nu-1}. \label{eqprf}
\end{align}
Note that the second term vanishes. The third term can be computed according to the form of $R$ and the condition \eqref{exPhinumu}, namely,
\[
\Phi_1\p_2^{-1}\Phi_1^{-1} e^\beta \p_1^{-1} e^{-\beta}=\Phi_2\p_1^{-1}\Phi_2^{-1} \p_2^{-1}.
\]
In more details, whenever $\mu=2$ and $\nu=1$,
\begin{align*}
	&\left(\res_{\p_2} \left( R\p_1\Phi_2^{-1} P_2^k  -P_2^k R\p_1 \Phi_2^{-1} \right) \Phi_1\p_{2}^{-1}\Phi_1^{-1}\right)_{<0}\\
	=&\left(\res_{\p_2}  R\p_2^k\Phi_{2}^{-1}\p_2^{-1}e^{\beta}\p_1e^{-\beta}
	- \res_{\p_2}  P_2^k R\Phi_{2}^{-1}\p_2^{-1}e^{\beta}\p_1e^{-\beta}\right)_{<0}=0;
\end{align*}
whenever $\mu=1$ and $\nu=2$,
\begin{align*}
	&\left(\res_{\p_1} \left( R\p_2\Phi_1^{-1} P_1^k  -P_1^k R\p_2 \Phi_1^{-1} \right) \Phi_2\p_{1}^{-1}\Phi_2^{-1}\right)_{<1}\\
	=&\left(\res_{\p_1}  R\p_1^k\Phi_{1}^{-1}e^{\beta}\p_1^{-1}e^{-\beta}\p_2
	- \res_{\p_1}  P_1^k R\Phi_{1}^{-1}e^{\beta}\p_1^{-1}e^{-\beta}\p_2\right)_{<1}=0.
\end{align*}
Thus the third term in \eqref{eqprf} also vanishes.
\end{itemize}
Taking \eqref{exlhs3} and \eqref{eqprf} together we conclude that $\hbox{l.h.s.}=0$.

Therefore the lemma is proved.
\end{prf}

As a consequence of Lemma~\ref{thm-exBkt}, we have the following result.
\begin{lem}\label{thm-extcomm}
The flows \eqref{exPhinut1}--\eqref{exPhinut2} commute with each other. Namely, for any $\ld, \mu, \nu\in\set{1,2}$ and $k, l\in\Z_{\geq1}$, it holds that
\begin{equation*}\label{extcomm}
\left[\frac{\p}{\p t_{\mu,k} },\frac{\p}{\p t_{\nu,l} }\right]\Phi_\ld=0.
\end{equation*}
\end{lem}

Note $a_{1,1}=\res_{\p_1}\Phi_{1}$ in \eqref{exPhi12}, and denote
\begin{equation}\label{rho}
\rho=\p_2(a_{1,1}).
\end{equation}

\begin{prp}\label{thm-ext12varrho}
	For $k\in\Z_{\ge1}$, it holds that
\begin{align}\label{ext12varrho}
	\frac{\p \rho}{\p t_{1,k}}=-\p_2\Big(\res_{\p_1}P_1^k\Big),\quad \frac{\p \rho}{\p t_{2,k}}=\p_2\left(P_2^k \p_2^{-1}\right)_{\geq0}(\rho).
\end{align}
\end{prp}
\begin{prf}
The proposition follows directly from the equations \eqref{exPhinut1}--\eqref{exPhinut2} and the definition of $\rho$ in \eqref{rho}.
\end{prf}

\begin{lem}\label{thm-exH1H2}
The constraint \eqref{exPhinumu} implies
\begin{align}
&\rho=\p_1\p_2(\beta)+\p_1(a_{2,1}), \label{rho2} \\
&\p_2\la\Phi_1\ra\Phi_1^{-1}=e^{\beta}\p_1^{-1}e^{-\beta}\rho,\label{exH120}\\
&\p_1\la\Phi_2\ra\Phi_2^{-1}=\p_1(\beta)+\p_2^{-1}\cdot\p_1(a_{2,1}),\label{exH210}
\end{align}
where $a_{2,1}=e^{-\beta}\res_{\p_2}\Phi_{2}$ (see \eqref{exPhi12}). Moreover, each side of the equality \eqref{exPhinumu} is equal to the following operator
\begin{align}\label{exH}
H=\p_1\p_2-\p_1(\beta)\p_2-\rho.
\end{align}
\end{lem}

\begin{prf}
The constraint  \eqref{exPhinumu} can be recast to
\begin{equation}\label{eqH}
(\p_1-\p_1(\beta))\cdot (\p_2-\p_2\la\Phi_1\ra\Phi_1^{-1}) =\p_2\p_1-\p_2\cdot\p_1\la\Phi_2\ra\Phi_2^{-1}.
\end{equation}
One observes that the left hand side does not contain negative powers in $\p_2$, while the right hand side does not contain negative powers in $\p_1$, hence each of them contain only nonnegative powers in $\p_1$ and $\p_2$. So we arrive at
\begin{align*}
&(\p_1-\p_1(\beta))(-\p_2\la\Phi_1\ra\Phi_1^{-1})
=\left((\p_1-\p_1(\beta))(-\p_2\la\Phi_1\ra\Phi_1^{-1})\right)_{\ge0}=-\p_2(a_{1,1})=-\rho, \\
&-\p_2\cdot\p_1\la\Phi_2\ra\Phi_2^{-1}=\left(-\p_2\cdot\p_1\la\Phi_2\ra\Phi_2^{-1}\right)_{\ge0} =-\p_2\cdot\p_1(\beta)-\p_1(a_{2,1}).
\end{align*}
Thus we conclude \eqref{exH120}--\eqref{exH}. What is more, substituting them into \eqref{eqH}, one has
\[
\p_1\p_2-\p_1(\beta)\p_2-\rho = \p_1\p_2-\p_2\cdot\p_1(\beta)-\p_1(a_{2,1}),
\]
which implies  the equality \eqref{rho2}. The lemma is proved.
\end{prf}

Now we are ready to show that the constraint \eqref{exPhinumu} is invariant with respect to the flows \eqref{exPhinut1}--\eqref{exPhinut2}, which is implied by the following lemma.
\begin{lem}\label{thm-ext12H1H2}
Suppose that the left hand side and the right hand side of \eqref{exPhinumu} are denoted respectively by
\begin{equation*}\label{exH12}
H_1=e^{\beta}\p_1e^{-\beta}\Phi_1\p_2\Phi_1^{-1},\quad H_2=\p_2\Phi_2\p_1\Phi_2^{-1}.
\end{equation*}
Then for $\nu,\mu\in\{1,2\}$ and $k\in\Z_{\geq1}$, the following equalities hold true:
\begin{align}\label{ext12H1H2}
\left(\frac{\p H_\mu}{\p t_{\nu,k}}\right)_{-}=0.
\end{align}
\end{lem}
\begin{prf}
Let us verify the equalities \eqref{ext12H1H2} case by case.
\begin{itemize}
\item[(i)] When $\nu=\mu=1$, according to \eqref{exPhi12} and \eqref{exPhinut2} one has
\begin{align*}
\frac{\p}{\p t_{1,k}}\Big(\p_1(\beta)\Big)=&\p_1\Big(e^{-\beta}\frac{\p e^\beta}{\p t_{1,k}}\Big) 
=\p_1\Big(e^{-\beta}( P_1^k)_{\geq0}(e^{\beta})\Big).
\end{align*}
With the help of \eqref{eqH} and \eqref{exH120}, we have
\begin{align*}
	\left(\frac{\p H_1}{\p t_{1,k}}\right)_{-}
=&\Bigg(\frac{\p}{\p t_{1,k}}\Bigg(\Big(\p_1-\p_1(\beta)\Big)\Big(\p_2-\p_2\la\Phi_1\ra\Phi_1^{-1} \Big) \Bigg)\Bigg)_-\\
	=&\Bigg(\p_1\Big(e^{-\beta}( P_1^k)_{\geq0}(e^{\beta})\Big)e^{\beta}\p_1^{-1}e^{-\beta}\rho \\
&+e^{\beta}\p_1e^{-\beta}\Big(\p_2\la( P_1^k)_{<0} \Phi_1\ra\Phi_1^{-1}-\p_2\la\Phi_1\ra\Phi_1^{-1}( P_1^k)_{<0}\Big)\Bigg)_-\\
	=&\Bigg(\p_1\Big(e^{-\beta}( P_1^k)_{\geq0}(e^{\beta})\Big) e^{\beta}\p_1^{-1}e^{-\beta}\rho +e^{\beta}\p_1e^{-\beta}\p_2\la( P_1^k)_{<0}\ra\\
	&+e^{\beta}\p_1e^{-\beta}(P_1^k)_{<0}\p_2\la \Phi_1\ra\Phi_1^{-1}-e^{\beta}\p_1e^{-\beta}\p_2\la\Phi_1\ra\Phi_1^{-1}( P_1^k)_{<0}\Bigg)_-\\
=&\Bigg(e^{\beta}\p_1\Big(e^{-\beta}( P_1^k)_{\geq0}(e^{\beta})\Big)\p_1^{-1}e^{-\beta}\rho +e^{\beta}\p_1e^{-\beta}\p_2\la P_1^k\ra \\
&+e^{\beta}\p_1e^{-\beta}\Big(P_1^k-(P_1^k)_{\geq0}\Big)\p_2\la \Phi_1\ra\Phi_1^{-1} \\
&-e^{\beta}\p_1e^{-\beta}\p_2\la\Phi_1\ra\Phi_1^{-1} \Big(P_1^k-(P_1^k)_{\geq0}\Big) \Bigg)_-\\
=&\Bigg(e^{\beta} \left[\p_1,\, e^{-\beta}( P_1^k)_{\geq0}(e^{\beta})\right] \p_1^{-1}e^{-\beta}\rho -e^{\beta}\p_1e^{-\beta}(P_1^k)_{\geq0}e^{\beta}\p_1^{-1} e^{-\beta}\rho\\
	&
	+e^{\beta}\p_1e^{-\beta}[\p_2,\, P_1^k] -e^{\beta}\p_1e^{-\beta}[\p_2\la\Phi_1\ra\Phi_1^{-1},\,P_1^k] +\rho (P_1^k)_{\geq0} \Bigg)_-\\
=&\Bigg(-e^{\beta} e^{-\beta}(P_1^k)_{\geq0}(e^{\beta})\p_1\cdot \p_1^{-1} e^{-\beta}\rho + e^{\beta}\p_1e^{-\beta}\left[\Phi_1\p_2\Phi_1^{-1}, \,\Phi_1\p_1^k\Phi_1^{-1}\right] \Bigg)_- \\
=&0.
\end{align*}

\item[(ii)] When $\nu=\mu=2$, by using \eqref{eqH}, \eqref{exPhinut1} and \eqref{exH210} we have
\begin{align*}
\left(\frac{\p H_2}{\p t_{2,k}}\right)_{-}=&-\Big(\frac{\p }{\p t_{2,k}}(\p_2\cdot\p_1\la\Phi_2\ra\Phi_2^{-1})\Big)_-\\
=&-\Big(-\p_2\cdot \p_1\la(P_2^k)_{<1}\Phi_2\ra\Phi_2^{-1}+\p_2\cdot\p_1\la\Phi_2\ra\Phi_2^{-1}(P_2^k)_{<1}\Big)_-\\
=&\Big(\p_2\cdot \p_1\la(P_2^k)_{<1}\ra+\p_2(P_2^k)_{<1}\cdot \p_1\la\Phi_2\ra\Phi_2^{-1} -\p_2\cdot\p_1\la\Phi_2\ra\Phi_2^{-1}(P_2^k)_{<1}\Big)_-\\
=&\Bigg(\p_2\cdot \p_1\la P_2^k\ra+\p_2(P_2^k)_{<1}\cdot \Big(\p_1(\beta)+\p_2^{-1}\cdot\p_1(a_{2,1})\Big) \\
&-\p_2\Big(\p_1(\beta) +\p_2^{-1}\cdot\p_1(a_{2,1})\Big)(P_2^k)_{<1}\Bigg)_-\\
=&\Bigg(\p_2\cdot \p_1\la P_2^k\ra +\p_2P_2^k \Big(\p_1(\beta)+\p_2^{-1}\cdot\p_1(a_{2,1})\Big) \\
&-\p_2\Big(\p_1(\beta)+\p_2^{-1}\cdot\p_1(a_{2,1})\Big)P_2^k\Bigg)_-\\
=&\Bigg(\p_2[\p_1,\,P_2^k] -\p_2\Big[\p_1(\beta)+\p_2^{-1}\cdot\p_1(a_{2,1}),\,P_2^k\Big]\Bigg)_-
\\
=&\Bigg(\p_2\Big[\p_1-\p_1\la\Phi_2\ra\Phi_2^{-1},\,P_2^k\Big]\Bigg)_-\\
=&\Bigg(\p_2\Big[\Phi_2\p_1\Phi_2^{-1},\,\Phi_2\p_2^k\Phi_2^{-1}\Big]\Bigg)_-=0.
\end{align*}

\item[(iii)] When $\nu=1,\mu=2$, by using \eqref{exPhinut2} and \eqref{exH210}--\eqref{eqH} we have
\begin{align*}
\left(\frac{\p H_2}{\p t_{1,k}}\right)_{-}=&\left(-\frac{\p }{\p t_{1,k}}(\p_2\cdot\p_1\la\Phi_2\ra\Phi_2^{-1})\right)_-\\
=&\Big(-\p_2\cdot\p_1\la(P_1^k)_{\geq0}\la\Phi_2\ra\ra\Phi_2^{-1} +\p_2\cdot\p_1\la\Phi_2\ra\Phi_2^{-1}(P_1^k)_{\geq0}\la\Phi_2\ra\Phi_2^{-1}\Big)_-\\
=&\Bigg(-\res_{\p_1}\p_2\p_1(P_1^k)_{\geq0}\Phi_2\p_1^{-1}\Phi_2^{-1} \\
&+\Big(\p_2\cdot\p_1(\beta)+\p_1(a_{2,1})\Big) \res_{\p_1}(P_1^k)_{\geq0}\Phi_2\p_1^{-1}\Phi_2^{-1}\Bigg)_-\\
=&\Bigg(\res_{\p_1}\Big(-\p_2\p_1+\p_2\cdot\p_1(\beta)+\p_1(a_{2,1}) \Big)\left(P_1^k-(P_1^k)_{<0}\right)\Phi_2\p_1^{-1}\Phi_2^{-1}\Bigg)_-\\
=&\Bigg(\res_{\p_1}\Big(-\p_2\p_1+\p_2\cdot\p_1(\beta)+\p_1(a_{2,1}) \Big)P_1^k\Phi_2\p_1^{-1}\Phi_2^{-1}+\p_2\cdot\res_{\p_1}P_1^k\Bigg)_-\\
=&-\Big(\res_{\p_1} H P_1^k H^{-1}\p_2 \Big)_- \\
=&-\Big(\res_{\p_1}e^{\beta}\p_1e^{-\beta} \Phi_1\p_2\Phi_1^{-1}P_1^k\Phi_1\p_2^{-1}\Phi_1^{-1} e^{\beta}\p_1^{-1}e^{-\beta}\p_2\Big)_- \\
=&-\Big(\res_{\p_1}e^{\beta}\p_1e^{-\beta}P_1^ke^{\beta} \p_1^{-1}e^{-\beta}\p_2\Big)_- =0.
\end{align*}

\item[(iv)] When $\nu=2,\mu=1$, according to \eqref{exPhi12} and \eqref{exPhinut1} one has
\begin{align*}
\frac{\p\beta}{\p t_{2,k}}=e^{-\beta}\res_{\p_2}\left(\frac{\p\Phi_2}{\p t_{2,k}}\p_2^{-1}\right) =-e^{-\beta}\res_{\p_2}\left((P_2^k)_{<1}\Phi_2 \p_2^{-1}\right) = -\res_{\p_2}(P_2^k\p_2^{-1}).
\end{align*}
With the help of \eqref{eqH}, \eqref{exH120}, \eqref{exPhinut2} and \eqref{exH}, we obtain
\begin{align*}
	\left(\frac{\p H_1}{\p t_{2,k}}\right)_{-}
=&\left(\frac{\p }{\p t_{2,k}}\Big(\big(\p_1-\p_1(\beta)\big) (-\p_2\la\Phi_1\ra\Phi_1^{-1})\Big)\right)_-\\
	=&\bigg(-\p_1\Big(\res_{\p_2}(P_2^k\p_2^{-1}) \Big)\p_2\la\Phi_1\ra\Phi_1^{-1}) -e^{\beta}\p_1e^{-\beta}\p_2\la (P_2^{k})_{\geq1}\la\Phi_1\ra\ra\Phi_1^{-1}\\
	&+e^{\beta}\p_1e^{-\beta}\p_2\la\Phi_1\ra \Phi_1^{-1}(P_2^{k})_{\geq1}\la\Phi_1\ra\Phi_1^{-1}\bigg)_-\\
	=&\bigg(-\res_{\p_2}[\p_1,   P_2^k \p_2^{-1}] e^{\beta}\p_1^{-1}e^{-\beta}\rho \\
 &- e^{\beta}\p_1e^{-\beta}\res_{\p_2}\p_2 \left( P_2^k-(P_2^{k})_{<1}\right)\Phi_1\p_2^{-1}\Phi_1^{-1}\\
	&+e^{\beta}\p_1e^{-\beta}e^{\beta}\p_1^{-1}e^{-\beta}\rho \res_{\p_2}\left( P_2^k-(P_2^{k})_{<1}\right)\Phi_1\p_2^{-1}\Phi_1^{-1}\bigg)_-\\
	=&\bigg(-\res_{\p_2}[\p_1,\,P_2^k\p_2^{-1}]e^{\beta} \p_1^{-1}e^{-\beta}\rho -e^{\beta}\p_1e^{-\beta}\res_{\p_2}\p_2 P_2^{k}\Phi_1\p_2^{-1}\Phi_1^{-1} \\
&+e^{\beta}\p_1e^{-\beta}\Big(\res_{\p_2}P_2^k + \p_2\Big(\res_{\p_2}(P_2^k\p_2^{-1}) \Big)+  \res_{\p_2}(P_2^k\p_2^{-1})\cdot \p_2\la\Phi_1\ra\Phi_1^{-1}\Big) \\ &+\rho\Big(\res_{\p_2}P_2^k\Phi_1\p_2^{-1}\Phi_1^{-1} -\res_{\p_2}(P_2^k\p_2^{-1}) \Big)\bigg)_-\\
	=&\bigg(-\res_{\p_2}[\p_1,\,P_2^k\p_2^{-1}]e^{\beta} \p_1^{-1}e^{-\beta}\rho + \res_{\p_2}(e^{\beta}\p_1e^{-\beta}P_2^k\p_2^{-1})\cdot \p_2\la\Phi_1\ra\Phi_1^{-1} \\
	& -e^{\beta}\p_1e^{-\beta} \res_{\p_2}\p_2 P_2^{k}\Phi_1\p_2^{-1}\Phi_1^{-1}  + \rho\res_{\p_2}P_2^k\Phi_1\p_2^{-1}\Phi_1^{-1} \bigg)_-\\
	=&\bigg(-\res_{\p_2}[\p_1,\,e^\beta]e^{-\beta}P_2^k\p_2^{-1} e^{\beta}\p_1^{-1}e^{-\beta}\rho+\res_{\p_2} P_2^k\p_2^{-1} \p_1 e^{\beta}\p_1^{-1}e^{-\beta}\rho \\
	&-e^{\beta}\p_1e^{-\beta}\res_{\p_2}\p_2 P_2^{k}\Phi_2\p_1^{-1}\Phi_2^{-1}\p_2^{-1}e^{\beta}\p_1e^{-\beta}
\\
&+
\rho\res_{\p_2}P_2^k\Phi_2\p_1^{-1}\Phi_2^{-1}\p_2^{-1} e^{\beta}\p_1e^{-\beta}
\bigg)_-\\
	=&\bigg(\res_{\p_2} P_2^k\p_2^{-1} \cdot \left( - [\p_1,\,e^\beta]+ \p_1e^{\beta}\right)\p_1^{-1}e^{-\beta}\rho \\
	&-\res_{\p_2}e^{\beta}\p_1e^{-\beta}\p_2 \Phi_2\p_1^{-1}\Phi_2^{-1}\Phi_2\p_2^{k}\Phi_2^{-1}\p_2^{-1}e^{\beta}\p_1e^{-\beta}  \\
& +\rho\res_{\p_2}\Phi_2\p_1^{-1}\Phi_2^{-1}\Phi_2\p_2^{k}\Phi_2^{-1}\p_2^{-1} e^{\beta}\p_1e^{-\beta} \bigg)_-\\
	=&\left( \res_{\p_2} P_2^k\p_2^{-1} \cdot \rho -
\res_{\p_2}\Big(\p_1\p_2-\p_1(\beta)\p_2-\rho\Big) \Phi_2\p_1^{-1}\Phi_2^{-1}P_2^{k}\p_2^{-1}e^{\beta}\p_1e^{-\beta}\right)_-\\
	=&\left(- \res_{\p_2}\p_2 P_2^k\p_2^{-1}\cdot e^{\beta}\p_1e^{-\beta}\right)_-=0.
\end{align*}
\end{itemize}
Thus the lemma is proved.
\end{prf}

\begin{prfof}{Theorem~\ref{thm-welldef}}
	Taking Lemmas~\ref{thm-extcomm}, \ref{thm-exH1H2} and \ref{thm-ext12H1H2} together, we conclude the theorem.
\end{prfof}

Theorem~\ref{thm-welldef} means that equations \eqref{exPhinut1}--\eqref{exPhinumu} compose an integrable hierarchy.
In particular, if $\Phi_2=e^\beta$ is independent of $y=t_{2,1}$, then from  the constraint \eqref{exPhinumu} it follows that $\Phi_1$ is also independent of $y$, hence equations \eqref{exPhinut1} and \eqref{exPhinut2} become
\begin{align*}\label{}
\frac{\p \Phi_1}{\p t_{1, k}}&=-(P_1^k)_{<0}\Phi_1,\quad  \frac{\p e^{\beta}}{\p t_{1, k}}=(P_1^k)_{\geq0}(e^\beta), \quad \frac{\p \Phi_1}{\p t_{2, k}}=0, \quad
\frac{\p e^\beta}{\p t_{2, k}}=0.
\end{align*}
In other words, the hierarchy \eqref{exPhinut1}--\eqref{exPhinumu} is reduced to the KP hierarchy, together with a series of equations of its eigenfunction $e^\beta$.
On the other hand, if $\Phi_1=1$, then in \eqref{exPhinut1}--\eqref{exPhinut2} the equations of $\Phi_1$ is trivial and the equations of $\Phi_2$ become
\begin{align*}
 \frac{\p \Phi_2}{\p t_{1, k}}=\Phi_2,
\quad
\frac{\p \Phi_2}{\p t_{2, k}}=-(P_2^k)_{<1}\Phi_2.
\end{align*}
With $\Phi_2$ replaced by $e^{\sum_{k}t_{1,k}}\Phi_2$, one can eliminate the dependence on $t_{1,k}$ and make the constraint \eqref{exPhinumu} trivial, such that the hierarchy \eqref{exPhinut1}--\eqref{exPhinumu} is reduced to the mKP hierarchy \cite{JM1983}.

\begin{defn}
The system of equations \eqref{exPhinut1}--\eqref{exPhinumu} is called the \emph{KP-mKP hierarchy}.
\end{defn}

\begin{rmk}\label{rmk-red2BKP}
If we take $\beta=0$ and assume $\Phi_\nu^{*}=\p_\nu\Phi_\nu^{-1}\p_\nu^{-1}$ with $\nu\in\{1,2\}$, then the flows $\p/\p t_{\nu,k}$ in \eqref{exPhinut1}--\eqref{exPhinumu} with $k\in\Zop$ are well defined. Such reduced flows compose the two component BKP hierarchy; see \cite{GHW2023} and references therein.
\end{rmk}

Let us describe the KP-mKP hierarchy in a more explicit way.
From \eqref{exH} one has
\begin{align*}
H^{-1}=&\left( (1-\p_1(\beta)\p_1^{-1}-\rho\p_1^{-1}\p_2^{-1})\p_1\p_2\right)^{-1} \nn\\ =&\p_1^{-1}\p_2^{-1}\left(1+ \p_1(\beta)\p_1^{-1}+\rho\p_1^{-1}\p_2^{-1}+(\p_1(\beta)\p_1^{-1}+\rho\p_1^{-1}\p_2^{-1})^2+\dots\right).
\end{align*}
With this notation, the operators $B_{\nu,k}^\mu$ given in \eqref{exBk} with  $\nu\ne\mu$ are recast to
\begin{align*}
B_{1,k}^2=&\res_{\p_1}(P_1^k)_{\ge0}\Phi_2\p_1^{-1}\Phi_2^{-1} =\res_{\p_1}(P_1^k)_{\ge0}H^{-1}\p_2,  \\
B_{2,k}^1=&\res_{\p_2}(P_2^k)_{\ge1}\Phi_1\p_2^{-1}\Phi_1^{-1}=\res_{\p_2}(P_2^k)_{\ge1} H^{-1}e^{\beta}\p_1 e^{-\beta}.
\end{align*}
It implies that the coefficients of the operators on the right hand side of \eqref{exPnut} are differential polynomials in the functions of the set
\begin{equation}\label{urho}
\left\{u_{\nu,k},\,\rho,\,\beta\mid k\in\Z_{\geq1}; \  \nu=1,2 \right\},
\end{equation}
where $u_{\nu,k}$ are given in \eqref{exP2}. Hence it is defined by \eqref{exPnut} a system of evolutionary equations of the unknown functions in \eqref{urho}.

\begin{exa}
Denote $u=u_{1,1}$, $v=u_{2,1}$, and let the subscripts $x$ and $y$ stand for derivatives with respect to them, i.e., $f_x=\p_1(f)$, $f_{yy}=\p_2^2(f)$, etc. With the help of \eqref{u12}, \eqref{rho} and \eqref{ext12varrho}, it is straightforward to calculate:
\begin{align}
&u_{y}+\rho_x=0, \quad v_x+\rho_y=\beta_{xyy},\nn\\
&\frac{\p u}{\p t_{2,2}}=(2\rho\beta_y-\rho_y)_x,\nn \\
&\frac{\p u}{\p t_{1,3}}=3 u u_x+\frac{1}{4}u_{x x x}+\frac{3}{4}\int \frac{\p^2 u}{\p {t_{1,2}}^2}\od x, \label{ut13}\\
&\frac{\p u}{\p t_{2,3}}=\left(-3 v\rho-3\rho{\beta_y}^2+3\rho\beta_{y y}+3\rho_y\beta_y-\rho_{y y} \right)_x, \nn\\
&\frac{\p v}{\p t_{1,2}}=(-2\rho\beta_x-\rho_x+2\beta_x\beta_{x y}+\beta_{x x y})_y, \nn\\
&\frac{\p v}{\p t_{2,3}}=3 v v_y+\frac{1}{4}v_{y y y}+ \frac{3}{4}\int \frac{\p^2 v}{\p {t_{2,2}}^2}\od y,\nn\\
&\frac{\p v}{\p t_{1,3}}=\left( -3u\rho+3 u\beta_{x y}-3\rho{\beta_x}^2-3\rho\beta_{x x}-\rho_{x x}-3\rho_x\beta_x \right. \nn \\
&\qquad\qquad\qquad \left.+3{\beta_x}^2\beta_{x y}+3\beta_{x x}\beta_{x y}+3\beta_{x}\beta_{x x y}+\beta_{x x x y}\right)_y,\nn\\
&\frac{\p \beta}{\p t_{1,2}}=2 u + {\beta_x}^2+\beta_{x x}, \label{betat12}  \\
&\frac{\p \beta}{\p t_{2,2}}=-2 v-{\beta_y}^2+\beta_{y y}, \label{betat22} \\
&\frac{\p \beta}{\p t_{1,3}}=3 u\beta_x+ {\beta_x}^3+3\beta_x\beta_{x x}+\beta_{x x x}+\frac{3}{2}u_x+\frac{3}{2}\int\frac{\p u}{\p t_{1,2}}\od x,  \nn\\
&\frac{\p \beta}{\p t_{2,3}}=3 v\beta_y +{\beta_y}^3-3\beta_y\beta_{y y}+\beta_{y y y}-\frac{3}{2} v_y-\frac{3}{2}\int\frac{\p v}{\p t_{2,2}}\od y. \label{betat23}
\end{align}
Note that equation \eqref{ut13} is just the KP equation, and equation \eqref{betat23} together with equation \eqref{betat22} gives the modified KP equation \cite{JM1983} of $\varphi=\beta_y$, say,
\begin{equation*}\label{}
 \frac{\p\varphi}{\p t_{2,3}}=\frac{1}{4}\varphi_{y y y}-\frac{3}{2}\varphi^2\varphi_y  +\frac{3}{4}\int  \frac{\p^2\varphi}{\p {t_{2,2}}^2}  \od y -\frac{3}{2}\varphi_y \int\frac{\p\varphi}{\p t_{2,2}}\od y.
\end{equation*}
\end{exa}

\section{Properties of the KP-mKP hierarchy}

In this section let us layout some more properties of the KP-mKP hierarchy.
\begin{prp}\label{thm-invol}
The KP-mKP hierarchy \eqref{exPhinut1}--\eqref{exPhinumu} admits the following
B\"{a}cklund transformations:
\begin{equation*}\label{BT}
\iota~:~(\Phi_1,\Phi_2)\mapsto\left.\left(e^{-\beta}\Phi_2,e^{-\beta}\Phi_1\right) \right|_{(\bt_1,\bt_2)\mapsto(\bt_2,\bt_1);\, (\p_1,\p_2)\mapsto(\p_2,\p_1)}.
\end{equation*}
Moreover, these B\"{a}cklund transformations are involutions, say, $\iota^2=\mathrm{Id}$.
\end{prp}
\begin{prf}
Denote $\Psi_\nu=e^{-\beta}\Phi_{\nu}$ for $\nu\in\{1,2\}$.
Suppose that it is verified the following equalities:
\begin{align}
\frac{\p \Psi_2}{\p t_{2, k}}=-\left(\Psi_2\p_2^k\Psi_2^{-1}\right)_{<0}\Psi_2,&\quad
\frac{\p \Psi_1}{\p t_{1, k}}=-\left(\Psi_1\p_1^k\Psi_1^{-1}\right)_{<1}\Psi_1, \nn \\
\frac{\p \Psi_2}{\p t_{1, k}}=\left(\Psi_1\p_1^k\Psi_1^{-1}\right)_{\geq1}\la\Psi_2\ra,&\quad
 \frac{\p \Psi_1}{\p t_{2, k}}=\left(\Psi_2\p_2^k\Psi_2^{-1}\right)_{\geq0}\la\Psi_2\ra, \nn \\
e^{-\beta}\p_2 e^{\beta}\Psi_2\p_1\Psi_2^{-1}&=\p_1\Psi_1\p_2\Psi_{1}^{-1} \label{exPsinumu}
\end{align}
with $k\in\Z_{\ge1}$,
then the first assertion is confirmed by exchanging the time variables $\bt_1$ and $\bt_2$ as well as the derivatives with respect to them. In fact, the equality \eqref{exPsinumu} follows immediately from \eqref{exPhinumu}. Furthermore, by using \eqref{exPhinut1}--\eqref{exPhinut2}, it is straightforward to show
\begin{align*}
\frac{\p \Psi_1}{\p t_{1, k}}=&-e^{-\beta}\frac{\p e^\beta}{\p t_{1,k}}e^{-\beta}\Phi_1 +e^{-\beta}\frac{\p\Phi_1}{\p t_{1,k}} \\
=& -e^{-\beta}\left(\Phi_1\p_1^k\Phi_1^{-1}\right)_{\ge0}(e^{\beta})\cdot e^{-\beta}\Phi_1 -e^{-\beta}\left(\Phi_1\p_1^k\Phi_1^{-1}\right)_{<0}\Phi_1 \\
=& -\left(e^{-\beta}\Phi_1\p_1^k\Phi_1^{-1}e^{\beta}\right)_{<1}e^{-\beta}\Phi_1 \\
=&-\left(\Psi_1\p_1^k\Psi_1^{-1}\right)_{<1}\Psi_1, \\
\frac{\p \Psi_2}{\p t_{1, k}}=&-e^{-\beta}\frac{\p e^\beta}{\p t_{1,k}}e^{-\beta}\Phi_2 +e^{-\beta}\frac{\p\Phi_2}{\p t_{1,k}} \\
=& -e^{-\beta}\left(\Phi_1\p_1^k\Phi_1^{-1}\right)_{\ge0}(e^{\beta})\cdot e^{-\beta}\Phi_2 +e^{-\beta}\left(\Phi_1\p_1^k\Phi_1^{-1}\right)_{\ge0}\la\Phi_2\ra \\
=& -\res_{\p_1}\left(e^{-\beta} \Phi_1\p_1^k\Phi_1^{-1}e^{\beta}\p_1^{-1}\right)\cdot e^{-\beta}\Phi_2 +\left(e^{-\beta}\Phi_1\p_1^k\Phi_1^{-1}e^{\beta}\right)_{\ge0}\la e^{-\beta}\Phi_2\ra \\
=&\left(\Psi_1\p_1^k\Psi_1^{-1}\right)_{\geq1}\la\Psi_2\ra,
\end{align*}
and the cases $\p\Phi_\nu/\p t_{2,k}$ are similar. So the first assertion is verified.
The second assertion is clear. Therefore the proposition is proved.
\end{prf}

As an application of the proposition, one can derive the Miura transformations between the KP hierarchy and the modified KP hierarchy studied in \cite{OWRC1993}. More exactly, suppose that $\Phi_1$ of the form \eqref{exPhi12} is the dressing operator of the KP hierarchy and that $e^{\beta}$ is an eigenfunction of it, then $e^{-\beta}\Phi_1$ is a dressing operator that solves the modified KP hierarchy. Conversely,  suppose that $\Phi_2$ of the form \eqref{exPhi12} is the dressing operator of the modified KP hierarchy, then $e^{-\beta}\Phi_2$ is a dressing operator of the KP hierarchy and $e^{\beta}$ is an eigenfunction of it.

The following result will be applied in the forthcoming section.
\begin{prp}\label{thm-exHequi}
The constraint \eqref{exPhinumu} yields the following equality
\begin{align}\label{exHequi}
\Phi_1\p_2^{-1}\Phi_1^{-1} =\p_2^{-1}+\Phi_2\p_1^{-1}\Phi_{2}^{-1}\p_2^{-1}\rho\p_2^{-1}.
\end{align}
Conversely, if $\rho\ne0$, then the equality \eqref{exHequi} implies the constraint \eqref{exPhinumu}.
\end{prp}
\begin{prf}
According to \eqref{exPhinumu} and \eqref{exH}, we have
\begin{align*}
\Phi_1\p_2^{-1}\Phi_1^{-1}&=\Phi_2\p_1^{-1}\Phi_2^{-1}\p_2^{-1}e^{\beta}\p_1e^{-\beta}\\
&=\Phi_2\p_1^{-1}\Phi_2^{-1}\p_2^{-1}\left(\p_1-\p_1(\beta)\right)\\
&=\Phi_2\p_1^{-1}\Phi_2^{-1}\p_2^{-1}(H+\rho)\p_2^{-1}\\
&=\Phi_2\p_1^{-1}\Phi_2^{-1}\p_2^{-1}\left(\p_2\Phi_2\p_1\Phi_2^{-1}+\rho \right)\p_2^{-1}\\
&=\p_2^{-1}+\Phi_2\p_1^{-1}\Phi_{2}^{-1}\p_2^{-1}\rho\p_2^{-1}.
\end{align*}

Conversely, since $\rho\ne0$, then from the equality \eqref{exHequi} one has
\begin{align*}
	\Phi_2\p_1^{-1}\Phi_2^{-1}\p_2^{-1} 
	&=\Big(\Phi_1\p_2^{-1}\Phi_1^{-1}\p_2-\Phi_1\p_2^{-1}\p_2\Phi_1^{-1}\Big)\rho^{-1}\\
	&=-\Phi_1\p_2^{-1}\cdot\p_2\la\Phi_1^{-1}\ra\rho^{-1}\\
	&=\Phi_1\p_2^{-1}\Phi_1^{-1}\p_2\la\Phi_1\ra\Phi_1^{-1}\rho^{-1},
\end{align*}	
namely,
\begin{equation}\label{eqH2}
	\p_2\Phi_2\p_1\Phi_2^{-1}=\rho\Phi_1\left(\p_2\la\Phi_1\ra\right)^{-1}\Phi_1\p_2\Phi_1^{-1}.
\end{equation}
It is easy to see that the operator $\left(\p_2\la\Phi_1\ra\right)^{-1}$ takes the form
\[
\left(\p_2\la\Phi_1\ra\right)^{-1}=\frac{1}{\rho}\p_1+\sum_{i\ge0}g_i\p_1^{-i}.
\]
Moreover, one observes that the left side of \eqref{eqH2} contains no nonnegative powers in $\p_1$, while the right side contains no nonnegative powers in $\p_2$, hence both sides contain only nonnegative powers in $\p_1$ and $\p_2$. For this reason, we have
\begin{equation}\label{eqH3}
\p_2\Phi_2\p_1\Phi_2^{-1}
=\left(\p_2\Phi_2\p_1\Phi_2^{-1}\right)_+ =\p_1\p_2-\p_1(\beta)\p_2-\p_1\p_2(\beta)-\p_1(a_{2,1}),
\end{equation}
and
\begin{align}
\rho\Phi_1\left(\p_2\la\Phi_1\ra\right)^{-1}\Phi_1\p_2\Phi_1^{-1} =&\left( \rho\Phi_1\left(\p_2\la\Phi_1\ra\right)^{-1}\Phi_1\p_2\Phi_1^{-1}\right)_+
\nn\\
=& \left(  \rho\Phi_1\left(\p_2\la\Phi_1\ra\right)^{-1}\left(\p_2-\p_2\la\Phi_1\ra\Phi_1^{-1}\right) \right)_+  \nn\\
=& \left(  \rho\Phi_1\left(\p_2\la\Phi_1\ra\right)^{-1} \p_2- \rho \right)_+ \nn\\
=& \left(\rho\Phi_1\left(\p_2\la\Phi_1\ra\right)^{-1}\right)_+\p_2-\rho \nn\\
=&(\p_1+g)\p_2-\rho  \label{eqH4}
\end{align}
for some function $g$.
Taking \eqref{eqH2}--\eqref{eqH4} together, we obtain
\[
g=-\p_1(\beta), \quad \rho=\p_1\p_2(\beta)+\p_1(a_{2,1}), \quad  \rho\Phi_1\Big(\p_2\la\Phi_1\ra\Big)^{-1}=\p_1-\p_1(\beta) .
\]
Thus the equality \eqref{eqH2} is recast to
\[
\p_2\Phi_2\p_1\Phi_2^{-1} =\Big(\p_1-\p_1(\beta)\Big)\Phi_1\p_2\Phi_1^{-1}=e^{\beta}\p_1e^{-\beta}\Phi_1\p_2\Phi_1^{-1},
\]
which is just \eqref{exPhinumu}.

Therefore the proposition is proved.
\end{prf}

\section{Baker-Akhiezer functions and bilinear equations}

In this section we want to rewrite the KP-mKP hierarchy \eqref{exPhinut1}--\eqref{exPhinumu} to the form of bilinear equations, and consider its relationship with the extended KP hierarchy studied in \cite{LW2021,WZ2016}.

Let $z$ be a parameter. By convention we assign certain actions of pseudo-differential operators on exponential functions as follows (recall $t_{1,1}=x$ and $t_{2,1}=y$)
\begin{equation*}\label{}
\left(\sum_{i}f_i \p_\nu^i\right) (e^{z t_{\nu,1}   })
=\left(\sum_{i}f_i z^i\right) e^{z t_{\nu,1} },
\end{equation*}
of which the left hand side is usually written as $\left(\sum_{i}f_i \p_\nu^i\right) e^{z t_{\nu,1}}$ for short.
\begin{lem}[see, for example, \cite{DKJM-KPBKP}] \label{thm-exres}
Given $\nu\in\{1,2\}$, for any pseudo-differential operators $A, B\in\mathcal{E}$ that contain only powers in $\p_\nu$, the following equality holds true:
\begin{equation*}\label{res}
\res_{z}\left(A e^{z t_{\nu,1} }\cdot B^* e^{-z t_{\nu,1}}\right)=\res_{\p_\nu} (A B).
\end{equation*}
Here and below $\res_{z}\left(\sum_i f_i z^i\right)=f_{-1}$.
\end{lem}

Given a solution of the KP-mKP hierarchy \eqref{exPhinut1}--\eqref{exPhinumu}, let us introduce a pair of Baker-Akhiezer functions
\begin{equation}\label{exwavefnu}
	w_\nu=w_\nu(\bt_1, \bt_2; z)=\Phi_\nu e^{\xi(\bt_\nu;z)}, \quad \nu\in\{1,2\},
\end{equation}
where $\bt_\nu=(t_{\nu,1},t_{\nu,2},t_{\nu,3},\dots)$ and
\[
\xi(\bt_\nu; z)=\sum_{k\in\Z_{\geq1}}
t_{\nu,k} z^k.
\]
According to \eqref{exPhinut1} and \eqref{exPhinut2}, it is easy to verify the following equalities:
\begin{equation}\label{exwavefnut}
\frac{\p w_\mu}{\p t_{\nu,k}}=(P_\nu^k)_{\geq \nu-1}w_\mu,\quad \mu,\nu\in\set{1,2}; \ k\in\Z_{\geq1}.
\end{equation}
We also introduce a pair of adjoint Baker-Akhiezer functions
\begin{align}
	w_1^{*}&=w_1^{*}(\bt_1, \bt_2; z)=(\Phi_1^{-1})^{*} e^{-\xi(\bt_1;z)}, \label{exwaveadf11}\\
	w_2^{*}&=w_2^{*}(\bt_1, \bt_2; z)=\bar{\Phi}_2^{*} e^{-\xi(\bt_2;z)}, \label{exwaveadf12}
\end{align}
where
\begin{align}
	\bar{\Phi}_2=\p_2^2 \Phi_2^{-1} \p_2^{-1}\rho\p_2^{-1}
=\left(\rho-(\rho a_{2,1}+2\rho\p_2(\beta)-\p_2(\rho))\p_2^{-1}+\dots \right)e^{-\beta}. \label{exwidePhi2}
\end{align}

\begin{thm}\label{thm-exbl2}
The Baker-Akhiezer functions and the adjoint Baker-Akhiezer functions satisfy
the following bilinear equation
\begin{equation}\label{exble}
\res_z\left(  w_1(\bt_1, \bt_2; z)w_1^{*}(\bt_1', \bt_2'; z) \right)=\res_z \left(z^{-2}w_2(\bt_1, \bt_2; z)w_2^{*}(\bt_1', \bt_2'; z)\right),
\end{equation}
with arbitrary time variables $(\bt_1, \bt_2)$ and $(\bt_1', \bt_2')$. Conversely, suppose that four functions of the form
\begin{align}
&w_1(\bt_1, \bt_2; z)=\left(1+\sum_{i\ge1}a_{1,i}(\bt_1, \bt_2)z^{-i}\right) e^{\xi(\bt_1;z)}, \label{exwavef1}\\
&w_2(\bt_1, \bt_2; z)=e^{\beta(\bt_1, \bt_2) }\left(1+\sum_{i\ge1}a_{2,i}(\bt_1, \bt_2)z^{-i}\right) e^{\xi(\bt_2;z)}, \label{exwavef2}\\
&w_1^{*}(\bt_1, \bt_2; z)=\left(1+\sum_{i\ge1}\tilde{a}_{1,i}(\bt_1, \bt_2)z^{-i}\right) e^{-\xi(\bt_1;z)}, \label{exwaveadf1}\\
&w_2^{*}(\bt_1, \bt_2; z)=e^{-\beta(\bt_1, \bt_2)  }\left(\rho(\bt_1, \bt_2)+\sum_{i\ge1}\tilde{a}_{2,i}(\bt_1, \bt_2)z^{-i}\right) e^{-\xi(\bt_2;z)}, \label{exwaveadf2}
\end{align}
with ${\p a_{1,1}}/{\p t_{2,1}}\ne0$,
satisfy the bilinear equation \eqref{exble}, then they are the Baker-Akhiezer functions and the adjoint Baker-Akhiezer functions of the
KP-mKP hierarchy \eqref{exPhinut1}--\eqref{exPhinumu}.
\end{thm}

\begin{prf}
The proof is similar with that of Theorem~3.9 in \cite{GHW2023}. Firstly, we  introduce the following set of indices
\[
\mathcal{I}=\left\{  (m_1,m_2,m_3,\dots )\mid m_i\in\Z_{\ge0} \hbox{ such that } m_i=0 \hbox{ for } i\gg 0\right\}.
\]
For $\bs{m}=(m_1,m_2,m_3,\dots )\in \mathcal{I}$, denote
\begin{equation}\label{dtm}
{\p_{\bt_\nu}}^{\bs m}=\prod_{k\ge1}\left(\frac{\p}{\p {t_{\nu,k}} }\right)^{m_k}, \quad \nu\in\{1,2\}.
\end{equation}
In order to show the equality \eqref{exble}, it suffices to verify
\begin{align}
& \res_{z}\Bigg({\p_{\bt_1}}^{\bs m}{\p_{\bt_2}}^{\bs n}\Big(w_1(\bt_1, \bt_2; z)\Big)\cdot w_1^*(\bt_1, \bt_2; z)\Bigg) \nn\\
=&\res_{z}\Bigg(z^{-2}{\p_{\bt_1}}^{\bs m}{\p_{\bt_2}}^{\bs n}\Big(w_2(\bt_1, \bt_2; z)\Big)\cdot w_2^*(\bt_1, \bt_2; z)\Bigg) \label{bl-exderiv2}
\end{align}
for any indices ${\bs m}, {\bs n}\in \mathcal{I}$. In fact, from \eqref{exwavefnut} one sees that there exists an operator $A^{\bs{m}, \bs{n}}\in \cE$ such that $A^{\bs{m}, \bs{n} }=(A^{\bs{m}, \bs{n} })_+$ and the following two equalities with $\nu\in\set{1,2}$ hold simultaneously:
\[
{\p_{\bt_1}}^{\bs m}{\p_{\bt_2}}^{\bs n}(w_\nu(\bt_1, \bt_2; z))=A^{\bs{m}, \bs{n}}(w_\nu(\bt_1, \bt_2; z)).
\]
By using \eqref{exwavefnu}, \eqref{exwaveadf11}, \eqref{exwaveadf12} and Lemma~\ref{thm-exres}, the equality \eqref{bl-exderiv2} is equivalent to
\begin{equation}\label{exresAmn}
\res_{\p_1}\left( \res_{\p_2}\left( A^{\bs{m}, \bs{n}}\Phi_1\p_2^{-1}\right)\cdot \Phi_1^{-1} \right)=
\res_{\p_2}\left( \res_{\p_1} \left( A^{\bs{m}, \bs{n}}\Phi_2\p_1^{-1} \p_2^{-2}\right)\cdot\bar{\Phi}_2\right).
\end{equation}
By using \eqref{exHequi} and \eqref{exwidePhi2} one has
\begin{align*}
\mathrm{l.h.s.}= &\res_{\p_1} \res_{\p_2} \left(A^{\bs{m}, \bs{n}}\Phi_1\p_2^{-1}\Phi_1^{-1} \right)\\
=& \res_{\p_2} \res_{\p_1} \left(A^{\bs{m}, \bs{n}}\left(\Phi_2\p_1^{-1}\Phi_2^{-1}\p_2^{-1}\rho\p_2^{-1}+\p_2^{-1} \right)\right) \\
=& \res_{\p_2} \res_{\p_1} \left(A^{\bs{m}, \bs{n}}\left(\Phi_2\p_1^{-1}\Phi_2^{-1}\p_2^{-1}\rho\p_2^{-1} \right)\right)=\mathrm{r.h.s.}
\end{align*}
This shows that the equality \eqref{exresAmn} is valid, and so is the equality \eqref{exble}.

Conversely, observe that the functions \eqref{exwavef1}--\eqref{exwaveadf2} can be uniquely represented in the form:
\begin{align*}
w_\nu(\bt_1, \bt_2; z)=\Phi_\nu e^{\xi(\bt_\nu;z)},\quad
w_\nu^{*}(\bt_1, \bt_2; z)=\widetilde{\Phi}_\nu^* e^{-\xi(\bt_\nu;z)},
\end{align*}
where $\Phi_\nu$ are pseudo-differential operators of the form \eqref{exPhi12}, and
\[
\widetilde{\Phi}_1=1+\sum_{i\ge1}\p_1^{-i} \tilde{a}_{1,i},  \quad \widetilde{\Phi}_2=\left(\rho+\sum_{i\ge1}\p_2^{-i}\tilde{a}_{2,i} \right)e^{-\beta}.
\]
Based on the bilinear equation \eqref{exble}, we obtain the following results.
\begin{itemize}
\item[(i)] For any $i\in\Z_{\ge0}$, let $\p_1^{i}$ act on \eqref{exble} and let $(\bt_1', \bt_2')=(\bt_1, \bt_2)$, then according to Lemma~\ref{thm-exres} one derives
\begin{equation*}\label{ }
	\res_{\p_1 }\left(  \p_1^{i}\Phi_1\widetilde{\Phi}_1  \right)=\res_{\p_2 } \left(\p_1^{i}\la{\Phi_2 }\ra \p_2^{-2}\widetilde{\Phi}_2 \right).
\end{equation*}
Clearly, the right hand side vanishes, hence $\Phi_1\widetilde{\Phi}_1 =\left(\Phi_1\widetilde{\Phi}_1 \right)_+=1$, namely,
\begin{equation*}\label{exPhi1wan}
\widetilde{\Phi}_1= \Phi_1^{-1} .
\end{equation*}
\item[(ii)]
For any $i,j\in\Z_{\ge0}$, let $\p_1^{i}\p_2^{j}$ act on \eqref{exble} and let $(\bt_1', \bt_2')=(\bt_1, \bt_2)$, then by using Lemma \ref{thm-exres} we have
\[
\res_{\p_1 } \Big(\p_1^{i}\p_2^{j}\la\Phi_1\ra \Phi_1^{-1}\Big)
	=\res_{\p_2 } \left(  \p_2^{j}\p_1^{i}\la\Phi_2\ra \p_2^{-2}\widetilde{\Phi}_2  \right).
\]
Namely, we obtain
\begin{equation}\label{blij}
	\res_{\p_1 }\res_{\p_2 } \Big(\p_1^{i}\p_2^{j}\Phi_1 \p_2^{-1}\Phi_1^{-1}\Big)
	=\res_{\p_2 }\res_{\p_1 }\left(  \p_1^{i}\p_2^{j}\Phi_2\p_1^{-1}\p_2^{-2}\widetilde{\Phi}_2  \right).
\end{equation}
It is easy to see the following expansions:
\[
\Phi_1 \p_2^{-1}\Phi_1^{-1}=\p_2^{-1}+\sum_{k\ge1}\sum_{l\ge2} f_{k l}\p_1^{-k}\p_2^{-l}, \quad \Phi_2\p_1^{-1}\p_2^{-2}\widetilde{\Phi}_2 =\sum_{k\ge1}\sum_{l\ge2}  g_{k l}\p_1^{-k}\p_2^{-l}
\]
with certain coefficients $f_{k l}$ and $g_{k l}$.
Hence from \eqref{blij} it follows that
\begin{equation}\label{exH3}
\Phi_1 \p_2^{-1}\Phi_1^{-1}-\p_2^{-1}
	=\Phi_2\p_1^{-1}\p_2^{-2}\widetilde{\Phi}_2 ,
\end{equation}
namely,
\[
	\p_1\Phi_2^{-1}\p_{2}^{-1}\p_2\la\Phi_1\ra \p_2^{-1} \Phi_1^{-1}=\p_2^{-2}\widetilde{\Phi}_2 .
\]
Note that the right hand side does not depend on $\p_1$, and that on the left hand side
the terms independent of $\p_1$ are $\Phi_2^{-1}\p_{2}^{-1}\cdot\p_2(a_{1,1})\cdot\p_2^{-1}$. Thus, with $\rho=\p_2(a_{1,1})$ we arrive at
\begin{equation*}
	\widetilde{\Phi}_2 =\p_2^{2}\Phi_2^{-1}\p_2^{-1}\rho\p_2^{-1}. \label{exPhi2wan2}
\end{equation*}
Substituting it into \eqref{exH3}, we obtain
\begin{equation}\label{exPhi1Phi212}
\Phi_1 \p_2^{-1}\Phi_1^{-1}-\p_2^{-1}
	=\Phi_2\p_1^{-1}\Phi_2^{-1}\p_2^{-1}\rho\p_2^{-1},
\end{equation}
which is equivalent to the constraint \eqref{exPhinumu} due to $\rho\ne0$ and Proposition~\ref{thm-exHequi}.
\item[(iii)]
For any $i\in\Z_{\ge0}$ and $k\in\Z_{\ge1}$, let $\p_1^{i}\frac{\p}{\p t_{1,k} }$  act on both sides of \eqref{exble}, and take $(\bt_1', \bt_2')=(\bt_1, \bt_2)$, then the right hand side vanishes and we arrive at
\begin{align*}
\res_z\left(\p_1^{i}\Big(\frac{\p\Phi_1}{\p t_{1,k}}+\Phi_1\p_1^{k}\Big)e^{\xi(\bt_1;z)}\cdot (\Phi_1^{-1})^{*} e^{-\xi(\bt_1;z)}\right)
=0.
\end{align*}
That is, due to Lemma~\ref{thm-exres},
\begin{align*}
	\res_{\p_1}\left(\p_1^{i}\Big(\frac{\p\Phi_1}{\p t_{1,k}}+\Phi_1\p_1^{k}\Big) \Phi_1^{-1} \right)
	=0.
\end{align*}
Hence
\[
\left(\Big(\frac{\p\Phi_1}{\p t_{1,k}}+\Phi_1\p_1^{k}\Big) \Phi_1^{-1} \right)_{<0}=0,
\]
which leads to
\begin{align*}
	\frac{\p\Phi_1}{\p t_{1,k}}=-\left(\Phi_1\p_1^{k}\Phi_1^{-1}\right)_{<0}\Phi_1	.
\end{align*}
\item[(iv)] For any $i\in\Z_{\ge0}$ and $k\in\Z_{\ge1}$, let $\p_2^{i}\frac{\p}{\p t_{2,k} }$  act on both sides of \eqref{exble}, and take $(\bt_1', \bt_2')=(\bt_1, \bt_2)$, then by using Lemma~\ref{thm-exres} we have
\[
	\res_{\p_1} \p_2^{i}\la\frac{\p \Phi_1}{\p t_{2,k}}\ra \Phi_1^{-1} =\res_{\p_2} \left(\p_2^{i}\left(\frac{\p\Phi_2}{\p t_{2,k}}+\Phi_2\p_2^{k}\right)\p_2^{-2}\cdot  \p_2^{2}\Phi_2^{-1}\p_2^{-1}\rho\p_2^{-1} \right).
\]
Note that the left hand side is just $\p_2^{i}\left(\frac{\p a_{1,1}}{\p t_{2,k}}\right)$, hence we obtain
\begin{equation*}\label{}
\left(\left(\frac{\p\Phi_2}{\p t_{2,k}}+\Phi_2\p_2^{k}\right)\Phi_2^{-1}\p_2^{-1}\rho\p_2^{-1} \right)_{<0}=\frac{\p a_{1,1}}{\p t_{2,k}}\p_2^{-1}.
\end{equation*}
It implies that
\[
\left(\left(\frac{\p\Phi_2}{\p t_{2,k}}+\Phi_2\p_2^{k}\right)\Phi_2^{-1}\right)_{<1}=0,
\]
namely,
\begin{align}\label{exphi2t2k}
\frac{\p\Phi_2}{\p t_{2,k}}=-\left(\Phi_2\p_2^{k}\Phi_2^{-1} \right)_{<1}\Phi_2.
\end{align}
\item[(v)]
For any $ i\in\Z_{\ge0}$ and $k\in\Z_{\ge1}$, let $\p_1^{i}\frac{\p}{\p t_{2,k} }$ act on both sides of \eqref{exble}, and take $(\bt_1',\bt_2')=(\bt_1,\bt_2)$, then with the same method as before we have
\begin{equation}\label{exphi1tak2}
	\res_{\p_1}\p_1^{i}\frac{\p\Phi_1}{\p t_{2,k}}\Phi_1^{-1}
 =\res_{\p_2} \left( \p_1^{i}\la\frac{\p\Phi_2}{\p t_{2,k}}+\Phi_2\p_2^{k}\ra\p_2^{-2}\cdot  \p_2^{2}\Phi_2^{-1}\p_2^{-1}\rho\p_2^{-1} \right).
\end{equation}
The right hand side is, thanks to \eqref{exPhi1Phi212} and \eqref{exphi2t2k},
\begin{align*}
\mathrm{r.h.s.}=&
\res_{\p_2} \res_{\p_1}\left( \p_1^i\left( \frac{\p \Phi_2}{\p t_{2,k}}+\Phi_2\p_2^k\right)\p_1^{-1}\Phi_2^{-1}\p_2^{-1}\rho\p_2^{-1} \right) \\
=&
\res_{\p_1} \res_{\p_2} \left( \p_1^i(\Phi_2\p_2^k\Phi_2^{-1})_{\geq1}\Phi_2\p_1^{-1}\Phi_2^{-1}\p_2^{-1}\rho\p_2^{-1} \right)\\
=&
\res_{\p_1} \res_{\p_2} \left( \p_1^i(\Phi_2\p_2^k\Phi_2^{-1})_{\geq1}\Big(\Phi_1 \p_2^{-1}\Phi_1^{-1}-\p_2^{-1}\Big) \right)\\
=&\res_{\p_1}\left(\p_1^i \cdot (\Phi_2\p_2^k\Phi_2^{-1})_{\geq1}\la\Phi_1\ra \Phi_1^{-1}\right).
\end{align*}
Substituting it into \eqref{exphi1tak2} we arrive at
\[
\frac{\p\Phi_1}{\p t_{2,k}}\Phi_1^{-1}= (\Phi_2\p_2^k\Phi_2^{-1})_{\geq1}\la\Phi_1\ra \Phi_1^{-1},
\]
which leads to
\[
\frac{\p\Phi_1}{\p t_{2,k}}=(\Phi_2\p_2^k\Phi_2^{-1})_{\geq1}\la\Phi_1\ra.
\]
Similarly, for any $ i\in\Z_{\ge0}$ and $k\in\Z_{\ge1}$, let $\p_2^{i}\frac{\p}{\p t_{1,k} }$ act on both sides of \eqref{exble}, and take $(\bt_1',\bt_2')=(\bt_1,\bt_2)$, then one shows
\[
\frac{\p\Phi_2}{\p t_{1,k}}=(\Phi_1\p_1^k\Phi_1^{-1})_{\geq0}\la\Phi_2\ra.
\]
\end{itemize}
Therefore the theorem is proved.
\end{prf}

Clearly, if one takes $\bt_2'=\bt_2$ in the bilinear equation \eqref{exble}, then it is reduced to the well-known bilinear equation of the KP hierarchy with Baker-Akhiezer function $w_1$ and the time variables $\bt_1$. As an application of Theorem~\ref{thm-exbl2}, let us study the relationship between the KP-mKP hierarchy and the extension of KP hierarchy studied in \cite{LW2021,WZ2016}, which is a subhierarchy of the dispersive universal Whitham hierarchy \cite{KIM1994,SB2008}.

Let us recall the extended KP hierarchy following the notations used in \cite{LW2021}. Consider two pseudo-differential operators of the form:
\begin{equation*} \label{exPhi}
\Phi=1+\sum_{i\ge 1}a_i  \p_1^{-i}, \quad \hat{\Phi}=e^{\ta}\left(1+\sum_{i\ge 1}b_i\p_1^{i}\right),
\end{equation*}
where $a_i, \, \ta, \, b_i$ belong to a certain graded algebra of smooth functions of $x$.
We refer the readers to \cite{LWZ2011} for details on pseudo-differential operators that contain infinitely many positive powers of a derivation.
The following evolutionary equations are well defined:
\begin{align}
&\frac{\p \Phi}{\p t_k}=- \left(\Phi \p_1^k\Phi^{-1}\right)_{<0}\Phi, \quad
\frac{\p \hat{\Phi}}{\p t_k}= \left(\Phi \p_1^k\Phi^{-1}\right)_{\ge0}\hat{\Phi} -\dt_{k1} \hat\Phi \p_1, \label{exppt1}\\
&\frac{\p \Phi}{\p \hat{t}_k}=-\left(\hat\Phi \p_1^k\hat\Phi^{-1}\right)_{<0}\Phi, \quad \frac{\p
\hat{\Phi}}{\p \hat{t}_k}=\left(\hat\Phi \p_1^k\hat\Phi^{-1}\right)_{\ge0}\hat{\Phi}, \label{exppt2}
\end{align}
where $k\in\Z_{\geq1}$. These equations compose the so-called extended KP hierarchy, which is a subhierarchy of the dispersive Whitham hierarchy associated to the Riemann sphere with two points $\infty$ and $\p_1(\theta)$; see \cite{LW2021,SB2008}. Similar as before, we take $t_1=x$.

Given a solution of the extended KP hierarchy \eqref{exppt1}--\eqref{exppt2}, the Baker-Akhiezer functions and the adjoint Baker-Akhiezer functions are defined by:
\begin{align*}
\psi(\bt, \hat{\bt}; z)=\Phi e^{\xi(\bm{t};z)}&, \quad
\hat{\psi}(\bm{t}, \hat{\bm{t}}; z)=\hat{\Phi}  e^{x
z-\xi(\hat{\bm{t}};z^{-1})},\\
\psi^\dag(\bm{t}, \hat{\bm{t}}; z)=(\Phi^{-1})^{*} e^{-\xi(\bm{t};z)}&, \quad
 \hat{\psi}^\dag(\bm{t}, \hat{\bm{t}}; z)=(\hat{\Phi}^{-1})^{*}  e^{-xz+\xi(\hat{\bm{t}};z^{-1})},
\end{align*}
where $\bt=(t_1,t_2,t_3,\dots)$ and $\hat{\bt}=(\hat{t}_1,\hat{t}_2,\hat{t}_3,\dots)$. Observe that these (adjoint) Baker-Akhiezer functions take the form:
\begin{align*}
&\psi(\bt, \hat{\bt}; z)=\left(1+\sum_{i\ge1}a_i(\bt, \hat{\bt})z^{-i}\right) e^{\xi(\bt;z)},  \\
&\hat{\psi}(\bt, \hat{\bt}; z)=e^{\ta(\bt, \hat{\bt}) }\left(1+\sum_{i\ge1}b_i(\bt, \hat{\bt})z^{i}\right) e^{x
z-\xi(\hat{\bt};z^{-1})},  \\
&\psi^\dag(\bt, \hat{\bt}; z)=\left(1+\sum_{i\ge1} {a}_i^\dag(\bt, \hat{\bt})z^{-i}\right)   e^{-\xi(\bt;z)}, \\
&\hat{\psi}^\dag(\bt, \hat{\bt}; z)=e^{-\ta(\bt, \hat{\bt}) }\left(\varrho(\bt,\hat{\bt})+\sum_{i\ge1}{b}_i^\dag(\bt, \hat{\bt})z^{i}\right)  e^{-x
z+\xi(\hat{\bt};z^{-1})}, 
\end{align*}
where
\begin{equation*} \label{exrho}
\varrho=e^{\ta}\left(\hat{\Phi}^{-1}\right)^{*}(1)\ne0.
\end{equation*}
\begin{thm}[\cite{LW2021}] \label{thm-exbl3}
The extended KP hierarchy \eqref{exppt1}--\eqref{exppt2} is equivalent to the following bilinear equation of (adjoint) Baker-Akhiezer functions
\begin{equation}
\res_z\left(  \psi(\mathbf{t},\hat{\mathbf{t}}; z)\psi^\dag(\mathbf{t}',
\hat{\mathbf{t}}'; z) \right)=\res_z \left(\hat{\psi}(\mathbf{t},
\hat{\mathbf{t}}; z) \hat{\psi}^\dag(\mathbf{t}',\hat{\mathbf{t}}'; z) \right),\label{ex-blexKP}
\end{equation}
with arbitrary time variables $(\mathbf{t},\hat{\mathbf{t}})$ and $(\mathbf{t}',
\hat{\mathbf{t}}')$.
\end{thm}

Taking Theorems~\ref{thm-exbl2} and \ref{thm-exbl3} together, it leads to the following result.
\begin{cor}\label{thm-equ}
The following statements hold true:
\begin{itemize}
\item[(i)] The bilinear equations \eqref{exble} and \eqref{ex-blexKP} are converted to each other by identifying the (adjoint) Baker-Akhiezer functions as
    	\begin{align}
w_1(\bt_1,\bt_2;z)=\psi(\bt_1,\bt_2;z),& \quad w_1^{*}(\bt_1,\bt_2;z)=\psi^{\dag}(\bt_1,\bt_2;z) ,\label{exidwave1}\\
		w_2(\bt_1,\bt_2;z)=\hat{\psi}(\bt_1,-\bt_2;z^{-1}),& \quad w_2^{*}(\bt_1,\bt_2;z)=\hat{\psi}^{\dag}(\bt_1,-\bt_2;z^{-1});\label{exidwave2}
	\end{align}
\item[(ii)] Whenever $\rho\ne0$, the KP-mKP hierarchy \eqref{exPhinut1}--\eqref{exPhinumu} is equivalent to the extended KP hierarchy \eqref{exppt1}--\eqref{exppt2}, whose solutions are related by
\begin{align}\label{Phi12equ}
\Phi_1=\left.\Phi\right|_{(\bt,\hat{\bt})\mapsto(\bt_1,\bt_2)}, \quad \Phi_2=\left.\hat{\Phi}\right|_{(\bt,\hat{\bt})\mapsto(\bt_1,-\bt_2);\, \p_1^i\mapsto(-\p_2)^{-i}}\cdot e^{-x \p_2^{-1}}.
\end{align}
\end{itemize}
\end{cor}
\begin{prf}
If we do the replacement $z\mapsto 1/z$ on the right hand side of the bilinear equation \eqref{exble}, and substitute into it with \eqref{exidwave1}--\eqref{exidwave2}, then we obtain the bilinear equation \eqref{ex-blexKP}, and vice versa. So the first item is verified. Accordingly, with the help of the definitions of the (adjoint) Baker-Akhiezer functions, we conclude the second item. The corollary is proved.
\end{prf}

\begin{rmk}
In particular, the formulae \eqref{exidwave1}--\eqref{exidwave2}, or equivalently \eqref{Phi12equ}, yield
\begin{equation*}
\beta(\bt_1,\bt_2)=\ta(\bt_1,-\bt_2), \quad \rho(\bt_1,\bt_2)=\varrho(\bt_1,-\bt_2). \label{betarho}
\end{equation*}
It implies that, the B\"{a}cklund transformation \eqref{BT} induces a certain transformation of a subhierarchy of the dispersive Whitham hierarchy associated to the Riemann sphere such that the marked points change as
\[
\big(\infty,\p_1(\beta(\bt_1,-\bt_2))\big)\mapsto \big(-\p_1(\beta(-\bt_2,\bt_1)),\infty\big).
\]
\end{rmk}

For the KP-mKP hierarchy \eqref{exPhinut1}--\eqref{exPhinumu}, there is another version of bilinear equation besides \eqref{exble} (or \eqref{ex-blexKP}). More exactly, given a solution of the KP-mKP hierarchy, let us introduce
\begin{equation}\label{exadwave2}
w_1^\dag(\bt_1,\bt_2;z) =\left(\p_1\Phi_1^{-1}e^{\beta}\p_1^{-1}e^{-\beta}\right)^*e^{-\xi(\bt_1;z)}, \quad
w_2^\dag(\bt_1,\bt_2;z) =\left(\p_2\Phi_2^{-1}\p_2^{-1}\right)^*e^{-\xi(\bt_2;z)}.
\end{equation}
\begin{thm}\label{thm-ble2}
The KP-mKP hierarchy \eqref{exPhinut1}--\eqref{exPhinumu} is equivalent to the following bilinear equation
\begin{equation}\label{exble3}
\res_z\left( z^{-1} w_1(\bt_1, \bt_2; z)w_1^{\dag}(\bt_1', \bt_2'; z) \right)=\res_z \left(z^{-1}w_2(\bt_1, \bt_2; z)w_2^{\dag}(\bt_1', \bt_2'; z)\right)
\end{equation}
with arbitrary time variables $(\bt_1, \bt_2)$ and $(\bt_1', \bt_2')$.
\end{thm}
\begin{prf}
The proof is almost the same as that of Theorem~\ref{thm-exbl2}.
\end{prf}

\begin{rmk}
Under the assumption given in Remark~\ref{rmk-red2BKP}, one has
\[
w_\nu^\dag(\bt_1,\bt_2;z)=w_\nu(\bt_1,\bt_2;-z), \quad \nu\in\{1,2\}.
\]
Hence from \eqref{exble3} it is reduced to the bilinear equation of (adjoint) Baker-Akhiezer functions for the two-component BKP hierarchy \cite{DJKM-KPtype}; see also \cite{LWZ2011,Shi}.
\end{rmk}

\section{Tau functions}
Following the approach in \cite{DKV2022,Dickey,GHW2023}, we are to introduce two tau functions of the KP-mKP hierarchy, such that the hierarchy can be recast to
the form of Hirota bilinear equations of tau functions.

For $\nu\in\{1,2\}$, let us consider the following shifting operators
\[
G_\nu(z)=\exp\left(-\sum_{k=1}^{\infty}\frac{1}{k z^k}\frac{\p}{\p t_{\nu,k}} \right).
\]
Clearly, one has
\[
G_\nu f(\bt_\nu)=f\left(\bt_\nu-[z^{-1}]\right) \quad \hbox{ with } \quad [z^{-1}]=\left(\frac{1}{z}, \frac{1}{2z^2}, \frac{1}{3z^3}, \dots\right).
\]
What is more, for generic parameters $z, \zeta$ and $\ve$, the conventions as follows are assumed:
\begin{align}
	G_\nu(\zeta)e^{-\xi(\bt_\nu;z)}=& e^{-\xi(\bt_\nu;z)} \zeta\sum_{m=0}^{\infty}\frac{z^m}{\zeta^{m+1}}, \label{Gexp1} \\
	G_\nu(\ve)G_\nu(\zeta)e^{-\xi(\bt_\nu;z)}=& e^{-\xi(\bt_\nu;z)}\frac{\ve\zeta}{z(\ve-\zeta)} \left(\zeta\sum_{m=0}^{\infty}\frac{z^m}{\zeta^{m+1}} -\ve\sum_{m=0}^{\infty}\frac{z^m}{\ve^{m+1}}\right). \label{Gexp2}
\end{align}

Given a solution of the KP-mKP hierarchy \eqref{exPhinut1}--\eqref{exPhinumu}, we recall its Baker-Akhiezer functions in \eqref{exwavefnu} and \eqref{exadwave2}, and introduce the following formal power series in $z^{-1}$:
\begin{align}\label{phiz}
\phi_\nu(z)=w_\nu(\bt_1,\bt_2;z)e^{-\xi(\bt_\nu;z)}, \quad \phi^\dag_\nu(z)=w^\dag_\nu(\bt_1,\bt_2;z)e^{\xi(\bt_\nu;z)}
\end{align}
with $\nu\in\{1,2\}$.
Here and below, without any confusion we do not write explicitly the dependence on the time variables $(\bt_1,\bt_2)$ to avoid lengthy notations. It is easy to see that these series take the form:
\begin{align*}
\phi_1(z)=1+\sum_{i\ge1}a_{1,i}  z^{-i},&\quad \phi_2(z)=e^{\beta}\Big(1+\sum_{i\ge1}a_{2,i}  z^{-i}\Big),\\
\phi_1^{\dag}(z)=1+\sum_{i\ge1}\hat{a}_{1,1}  z^{-i},&\quad \phi_2^{\dag}(z)=e^{-\beta}\Big(1+\sum_{i\ge1}\hat{a}_{2,i} z^{-i}\Big).
\end{align*}
\begin{lem}\label{thm-exphiG12}
The following equalities hold true:
\begin{align}
&\phi_1(z)G_1(z)\phi_1^\dag(z)=e^{\beta}G_1(z)e^{-\beta}, \label{exphi1G1}\\
&\phi_2(z)G_2(z)\phi_2^\dag(z)=1, \label{exphi2G2} \\
&\p_1\log\phi_1(z)=(1-G_1(z))a_{1,1},  \label{phi1a1} \\
&\p_2\log\phi_2(z)=\p_2(\beta)+(1-G_2(z))a_{2,1}. \label{phi2a1}
\end{align}
\end{lem}
\begin{prf}
Let us take $\bt_1'=\bt_1-[\zeta^{-1}]$ and $\bt_2'=\bt_2$ in the bilinear equation \eqref{exble3}, with the help of \eqref{Gexp1} we have
\begin{align*}
\res_z\left( z^{-1}\phi_1(z)G_1(\zeta)\phi_1^{\dag}(z) \zeta\sum_{m=0}^{\infty}\frac{z^m}{\zeta^{m+1}} \right) =e^{\beta}G_1(\zeta)e^{-\beta}.
\end{align*}
The left hand side is just $\phi_1(\zeta)G_1(\zeta)\phi_1^{\dag}(\zeta)$, hence the equality \eqref{exphi1G1} is verified. Similarly, the equality \eqref{exphi2G2} follows by letting $\bt_2'=\bt_2-[\zeta^{-1}]$ and $\bt_1'=\bt_1$ in the bilinear equation \eqref{exble3}.

Recalling the adjoint Baker-Akhiezer functions given in \eqref{exwaveadf11} and \eqref{exwaveadf12}, we introduce
\[
\phi_\nu^*(z)=w_\nu^*(\bt_1,\bt_2;z)e^{\xi(\bt_1;z)}, \quad \nu\in\{1,2\}.
\]
It is easy to see that these series take the form
\[
\phi_1^*(z)=1-a_{1,1}z^{-1}+\hbox{l.o.t.}, \quad \phi_2^*(z)=e^{-\beta}\left(\rho-(\rho a_{2,1}+2\rho\p_2(\beta)-\p_2(\rho) )z^{-1}+\hbox{l.o.t.}\right),
\]
where `l.o.t.' stands for lower-order terms of powers in $z^{-1}$.
With the same method as before, by letting $\bt_1'=\bt_1-[\zeta^{-1}]$ and $\bt_2'=\bt_2$ in the bilinear equation \eqref{exble}, one has
\begin{align*}
\res_z\left(\phi_1(z)G_1(\zeta)\phi_1^{*}(z) \zeta\sum_{m=0}^{\infty}\frac{z^m}{\zeta^{m+1}} \right) =0,
\end{align*}
which implies
\begin{equation}\label{phi1eq2}
\phi_1(\zeta)G_1(\zeta)\phi_1^{*}(\zeta)-1=0.
\end{equation}
Letting $\p_1$ act on both sides of \eqref{exble}, and taking $\bt_1'=\bt_1-[\zeta^{-1}]$ and $\bt_2'=\bt_2$, one has
\begin{align*}
\res_z\left( \left(\p_1\phi_1(z)+z\phi_1(z)\right)G_1(\zeta)\phi_1^{*}(z) \zeta\sum_{m=0}^{\infty}\frac{z^m}{\zeta^{m+1}} \right) =0,
\end{align*}
which leads to
\[
\p_1\phi_1(\zeta)\cdot G_1(\zeta)\phi_1^{*}(\zeta)+\zeta \left(\phi_1(\zeta)G_1(\zeta)\phi_1^{*}(\zeta) -1-a_{1,1}\zeta^{-1}+G_1(\zeta)a_{1,1}\zeta^{-1}\right)=0.
\]
This equality together with \eqref{phi1eq2} implies the equality \eqref{phi1a1}. Accordingly, from \eqref{phi1a1} we obtain
\[
\p_2\log\left(e^{-\beta}\phi_2(z)\right)=(1-G_2(z))a_{2,1},
\]
which is just \eqref{phi2a1}.
The lemma is proved.
\end{prf}

For $\nu\in\{1,2\}$, we consider the following differential operators:
\begin{equation*}\label{exNnu}
	N_\nu(z)=-\frac{\p}{\p z}+\sum_{k=1}^{\infty}\frac{1}{z^{k+1}}\frac{\p}{\p t_{\nu,k}}.
\end{equation*}
It is easy to see that, for any function $f$ independent of the parameter $z$,
\begin{equation}
N_\nu(z)G_\nu(z) f=0. \label{exNGf}
\end{equation}

\begin{lem}\label{thm-NNphi}
For $\mu,\nu\in\set{1,2}$, the following equalities hold true:
\begin{equation}\label{exNmuNnu}
N_\nu(\ve)N_\mu(\zeta)\log\phi_\mu(\zeta)=N_\mu(\zeta)N_\nu(\ve)\log\phi_\nu(\ve).
\end{equation}
\end{lem}
\begin{prf}
Let us verify the equalities \eqref{exNmuNnu} case by case.
\begin{itemize}
\item[(i)]	When $\mu\ne\nu$,
let $\bt_1'=\bt_1-[\zeta^{-1}]$ and $\bt_2'=\bt_2-[\ve^{-1}]$ in the bilinear equation \eqref{exble3}, then according to \eqref{Gexp1} we have
\begin{align*}
	&\res_z \left( z^{-1}\phi_1(z)G_1(\zeta)G_2(\ve)\phi_1^{\dag}(z) \zeta\sum_{m=0}^{\infty}\frac{z^m}{\zeta^{m+1}}\right) \\
=&	\res_z \left(z^{-1}\phi_2(z)G_1(\zeta)G_2(\ve)\phi_2^{\dag}(z) \ve\sum_{m=0}^{\infty}\frac{z^m}{\ve^{m+1}}\right).
\end{align*}
Namely,
\begin{align*}
\phi_1(\zeta)G_1(\zeta)G_2(\ve)\phi_1^{\dag}(\zeta)=	\phi_2(\ve)G_1(\zeta)G_2(\ve)\phi_2^{\dag}(\ve).
\end{align*}
By virtue of \eqref{exphi1G1} and \eqref{exphi2G2} in Lemma \ref{thm-exphiG12}, this equality is recast to
\begin{align*}
\phi_1(\zeta)G_2(\ve)\frac{e^{\beta}G_1(\zeta)e^{-\beta}}{\phi_1(\zeta)} =\phi_2(\ve)G_1(\zeta)\frac{1}{\phi_2(\ve)},
\end{align*}
or equivalently,
\begin{align*}
\left(1-G_2(\ve)\right)\log\phi_1(\zeta)+G_2(\ve)\left(1-G_1(\zeta)\right) \beta=\left(1-G_1(\zeta)\right)\log\phi_2(\ve).
\end{align*}
Let $N_1(\zeta)N_2(\ve)$ act on both sides, then with the help of \eqref{exNGf} we arrive at
\begin{equation*}
	N_1(\zeta)N_2(\ve)\log\phi_1(\zeta)=N_1(\zeta)N_2(\ve)\log\phi_2(\ve).
\end{equation*}

\item[(ii)] When $\mu=\nu=1$, let $\bt_1'=\bt_1-[\zeta^{-1}]-[\ve^{-1}]$ and $\bt_2'=\bt_2$ in the bilinear equation \eqref{exble}, then by using \eqref{Gexp2} we have
\begin{equation*}
	\res_z \left( \phi_1(z)G_1(\zeta)G_1(\ve)\phi_1^*(z) \frac{\ve\zeta}{z(\ve-\zeta)} \left(\zeta\sum_{m=0}^{\infty}\frac{z^m}{\zeta^{m+1}} -\ve\sum_{m=0}^{\infty}\frac{z^m}{\ve^{m+1}}\right)\right)=0.
\end{equation*}
It leads to
\[
\phi_1(\zeta)G_1(\zeta)G_1(\ve)\phi_1^*(\zeta) =\phi_1(\ve)G_1(\zeta)G_1(\ve)\phi_1^*(\ve),
\]
which together with \eqref{phi1eq2} implies
\[
\left(1-G_1(\ve)\right)\log\phi_1(\zeta)=\left(1-G_1(\zeta)\right)\log\phi_1(\ve).
\]
Letting $N_1(\ve)N_1(\zeta)$ act on both sides, one obtains
\begin{equation}\label{N1N1}
N_1(\ve)N_1(\zeta)\log\phi_1(\zeta)=N_1(\zeta)N_1(\ve)\log\phi_1(\ve).
\end{equation}

\item[(iii)] When $\mu=\nu=2$, by using Proposition~\ref{thm-invol} and the equality \eqref{N1N1} we have
\[
N_2(\ve)N_2(\zeta)\log\left(e^{-\beta}\phi_2(\zeta)\right) =N_2(\ve)N_2(\zeta)\log\left(e^{-\beta}\phi_2(\ve)\right).
\]
Hence we conclude
\begin{equation*}\label{exN2N2}
N_2(\ve)N_2(\zeta)\log\phi_2(\zeta) =N_2(\ve)N_2(\zeta)\log \phi_2(\ve).
\end{equation*}
\end{itemize}
Therefore the lemma is proved.
\end{prf}

\begin{rmk}
For the purpose of shortening the proofs of Lemmas~\ref{thm-exphiG12} and \ref{thm-NNphi}, both bilinear equations \eqref{exble} and \eqref{exble3} have been employed. In fact, one can prove these two lemmas based on only the bilinear equation \eqref{exble3}, in consideration of its second-order derivatives with respect to $x$ and $y$.
\end{rmk}

With the help of the series given in \eqref{phiz}, we introduce a class of functions of $(\bt_1,\bt_2)$ as
\begin{equation*}\label{exetak}
	\eta_{\nu,k}=-\res_z z^k N_\nu(z)\log\phi_\nu(z),
\end{equation*}
where $\nu\in\set{1,2}$ and $k\in\Z_{\ge1}$. For example, we have
\begin{align}\label{eta11}
  \eta_{1,1}=&-\res_z\left( z\left(-\frac{\p}{\p z}+z^{-2}\frac{\p}{\p t_{1,1}}+\hbox{l.o.t.}\right) \log\left(1+a_{1,1}z^{-1}+\hbox{l.o.t.}\right) \right)=-a_{1,1},
\\
\eta_{2,1}=&-\res_z\left( z\left(-\frac{\p}{\p z}+z^{-2}\frac{\p}{\p t_{2,1}}+\hbox{l.o.t.}\right) \log\left( e^\beta \left(1+a_{2,1}z^{-1}+\hbox{l.o.t.}\right) \right) \right)\nn\\
=&-\p_2(\beta)-a_{2,1}. \label{eta21}
\end{align}
\begin{lem}
The following $1$-form is closed:
	\begin{equation*}\label{exeta}
		\eta=\sum_{\nu=1}^2\sum_{k=1}^{\infty} \eta_{\nu,k}\od t_{\nu,k}.
	\end{equation*}
\end{lem}
\begin{prf}
For any $\nu,\mu\in\{1,2\}$ and $k,l\in\Z_{\ge1}$, it is straightforward to show
\begin{align*}
\frac{\p\eta_{\nu,k}}{\p t_{\mu,l}}=&\res_{\zeta}\zeta^l N_{\mu}(\zeta)\eta_{\nu,k} \\
=&-\res_{\zeta}\res_{\ve}\zeta^l \ve^k N_{\mu}(\zeta)N_{\nu}(\ve)\log\phi_\nu(\ve) \\
=&-\res_{\ve}\res_{\zeta}\zeta^l \ve^k N_{\nu}(\ve)N_{\mu}(\zeta)\log\phi_\mu(\zeta) \\
=&\res_{\ve}\ve^k N_{\nu}(\ve)\eta_{\mu,l} \\
=&\frac{\p\eta_{\mu,l}}{\p t_{\nu,k}},
\end{align*}
in which the third equality is due to Lemma~\ref{thm-NNphi}. The lemma is proved.
\end{prf}

According to the lemma, there locally exists a function $\tau_1=\tau_1(\bt_1,\bt_2)$ such that
\begin{equation*}\label{}
	\eta=\od \log\tau_1.
\end{equation*}
Note that $\log\tau_1$ is determined up to addition of a linear function of the time variables $(\bt_1,\bt_2)$. Moreover, we introduce $\tau_2=e^\beta\tau_1$.
\begin{dfn}
The functions $\tau_1$ and $\tau_2$ are called the tau functions of the KP-mKP hierarchy \eqref{exPhinut1}--\eqref{exPhinumu}.
\end{dfn}

\begin{prp}\label{thm-exphi12tau}
The tau functions satisfy
\begin{align}
	\phi_1(z)&=\frac{G_1(z)\tau_1}{\tau_1}, \quad
	\phi_1^{\dag}(z)=\frac{G_1^{-1}(z)\tau_2}{\tau_2},\label{exphitau}\\ \phi_2(z)&=\frac{G_2(z)\tau_2}{\tau_1},\quad \phi_2^{\dag}(z)=\frac{G_2^{-1}(z)\tau_1}{\tau_2}. \label{exphitau2}
\end{align}
Here note that $G_\nu^{-1}(z)=\exp\left(\sum_{k=1}^{\infty}\frac{1}{k z^k}\frac{\p}{\p t_{\nu,k}} \right)$.
\end{prp}
\begin{prf}
In combination of the equalities \eqref{phi1a1}, \eqref{eta11} and $\eta_{1,1}=\p_1\log\tau_1$, one has
\begin{equation*}\label{p1delta1}
  \p_1\left(\log\phi_1(z)-(G_1(z)-1)\log\tau_1 \right)=0.
\end{equation*}
Since $\log\phi_1(z)-(G_1(z)-1)\log\tau_1$ is a series in $z^{-1}$ whose coefficients can be represented as differential polynomials in $\{a_{1,k}\}_{k=1}^\infty$ without constant terms, then it vanishes indeed. Hence the first equality in \eqref{exphitau} is verified. Subsequently, with the help of \eqref{exphi1G1} we have
\[
\phi^\dag_1(z)=\frac{G_1^{-1}(z)e^\beta}{e^\beta}\frac{1}{G_1^{-1}(z)\phi_1(z)} =\frac{G_1^{-1}(z)e^\beta}{e^\beta}\frac{G_1^{-1}(z)\tau_1}{\tau_1} =\frac{G_1^{-1}(z)\tau_2}{\tau_2},
\]
which is just the second equality in \eqref{exphitau}.

On the other hand, by using the equalities \eqref{phi2a1}, \eqref{eta21} and $\eta_{2,1}=\p_2\log\tau_1$, one has
\begin{equation*}\label{p1delta1}
  \p_2\log\phi_2(z)=\p_2(\beta)+(G_2(z)-1)\left(\p_2\log\tau_1+\p_2(\beta)\right),
\end{equation*}
that is,
\[
\p_2\left(\log\phi_2(z)-(G_2(z)-1)\log\tau_1-G_2(z)\beta\right)=0.
\]
For the same reason as before, it follows that
\[
\log\phi_2(z)-(G_2(z)-1)\log\tau_1-G_2(z)\beta=0.
\]
Hence the first equality in \eqref{exphitau2} is verified by
\[
\phi_2(z)=\frac{G_2(z)(\tau_1 e^\beta)}{\tau_1}=\frac{G_2(z)\tau_2}{\tau_1}.
\]
Moreover, by using \eqref{exphi2G2} we obtain
\[
\phi^\dag_2(z)=\frac{1}{G_1^{-1}(z)\phi_2(z)}=\frac{G_2^{-1}(z)\tau_1}{\tau_2}.
\]
Therefore the proposition is proved.
\end{prf}

This proposition immediately leads to the main theorem of the present section.
\begin{thm}\label{thm-extauKPmKP}
The Baker-Akhiezer functions given by \eqref{exwavefnu} and the adjoint Baker-Akhiezer functions  by \eqref{exadwave2} can be represented via the tau functions as:
\begin{align*}
w_1(\bt_1,\bt_2;z)=&\frac{\tau_1(\bt_1-[z^{-1}],\bt_2)}{\tau_1(\bt_1,\bt_2)} e^{\xi(\bt_1;z)},\\ w_1^{\dag}(\bt_1,\bt_2;z)=&\frac{\tau_2(\bt_1+[z^{-1}],\bt_2)}{\tau_2(\bt_1,\bt_2)} e^{-\xi(\bt_1;z)},\\ w_2(\bt_1,\bt_2;z)=&\frac{\tau_2(\bt_1,\bt_2-[z^{-1}])}{\tau_1(\bt_1,\bt_2)} e^{\xi(\bt_2;z)},\\ w_2^{\dag}(\bt_1,\bt_2;z)=&\frac{\tau_1(\bt_1,\bt_2+[z^{-1}])}{\tau_2(\bt_1,\bt_2)}e^{-\xi(\bt_2;z)}. 
\end{align*}
So the bilinear equation \eqref{exble3} can be recast to
\begin{align}
&\res_z \left(z^{-1}\tau_1(\bt_1-[z^{-1}],\bt_2)\tau_2(\bt_1'+[z^{-1}],\bt_2')e^{\xi(\bt_1-\bt_1';z)}\right) \nonumber\\
=&\res_z\left( z^{-1}\tau_2(\bt_1,\bt_2-[z^{-1}])\tau_1(\bt_1',\bt_2'+[z^{-1}]) e^{\xi(\bt_2-\bt_2';z)}\right). \label{bletau}
\end{align}
\end{thm}

The bilinear equation \eqref{bletau} can be represented in the form of Hirota equations \cite{Hir}. To this end let us introduce some notations.
We recall the set $\mathcal{I}$ of indices given in the proof of Theorem~\ref{thm-exbl2}. For $\bs{m}=(m_1,m_2,\dots)\in\mathcal{I}$, one has the following notations:
\begin{equation*}\label{}
|\bs{m}|=\sum_{i\ge1}m_i, \quad ||\bs{m}||=\sum_{i\ge1}i m_i, \quad \bs{m}!=\prod_{i\ge1}m_i !.
\end{equation*}
Moreover, given two elements $\bs{m}=(m_1,m_2,m_3,\dots)$ and $\bs{l}=(l_1,l_2,l_3,\dots)$ of $\mathcal{I}$, we write $\bs{l}\le\bs{m}$ if
\[
\bs{m}-\bs{l}=(m_1-l_1,m_2-l_2,m_3-l_3,\dots)\in\mathcal{I}.
\]
For $f$ and $g$ being two functions of $(\bt_1,\bt_2)$, the Hirota operators $D_{\nu,k}$ are defined by
\begin{equation}\label{Dknu}
D_{\nu,k}f\cdot g= \left.\frac{\p }{\p s}\right|_{s=0} \left(\left.f\right|_{t_{\nu,k}\mapsto t_{\nu,k}+s} \left.g\right|_{t_{\nu,k}\mapsto t_{\nu,k}-s} \right),
\end{equation}
where $\nu\in\{1,2\}$ and $k\in\Z_{\ge1}$. Denote
\[
\bs{D}_\nu=\left(D_{\nu,1}, D_{\nu,2}, D_{\nu,3},\dots \right), \quad \widetilde{\bs{D}}_\nu=\left(D_{\nu,1}, \frac{1}{2}D_{\nu,2}, \frac{1}{3}D_{\nu,3},\dots \right),
\]
and, for $\bs{m}=(m_1,m_2,\dots)\in\mathcal{I}$ we write (cf. \eqref{dtm})
\[
{\bs{D_\nu}}^{\bs m}=\prod_{i\ge1}\left(D_{\nu,i}\right)^{m_i}.
\]
\begin{prp}
Let $p_k$ with $k\ge0$ be the Schur polynomials given by
\[
e^{\xi(\bt_\nu;z)}=\sum_{k}p_k(\bt_\nu)z^k.
\]
The bilinear equation \eqref{bletau} can be recast to the following system of Hirota equations: for any $\bs{m}, \bs{n}\in\mathcal{I}$,
\begin{align}
& \left( \sum_{\bs{l}\le\bs{m}} \frac{2^{|\bs{l}|}}{(\bs{m}-\bs{l})!\,\bs{l}!\, \bs{n}!} p_{||\bs{l}||}(-\widetilde{\bs{D}}_1){\bs{D}_1}^{\bs{m}-\bs{l}} {\bs{D}_2}^{\bs{n}}\right. \nn\\
&\qquad \left. - \sum_{\bs{l}\le\bs{n}} \frac{(-1)^{|\bs{m}|+|\bs{n}-\bs{l}|}\,2^{|\bs{l}|}}{\bs{m}!\,(\bs{n}-\bs{l})!\,\bs{l}!} {\bs{D}_1}^{\bs{m}}{\bs{D}_2}^{\bs{n}-\bs{l}} p_{||\bs{l}||}(\widetilde{\bs{D}}_2)
\right)\tau_1\cdot\tau_2=0. \label{hir}
\end{align}
\end{prp}
\begin{prf}
For $\nu\in\{1,2\}$, denote $\bm{s}_\nu=(s_{\nu,1},s_{\nu,2},s_{\nu,3},\dots)$. According to \eqref{Dknu} on has
\[
\exp\left(\sum_{\nu=1}^2\sum_{k\ge1}s_{\nu,k}D_{\nu,k}\right)f\cdot g=f(\bt_1+\bm{s}_1,\bt_2+\bm{s}_2)g(\bt_1-\bm{s}_1,\bt_2-\bm{s}_2).
\]
Let us do the replacements:
\[
\bt_\nu\mapsto\bt_\nu+\bm{s}_\nu, \quad \bt_\nu'\mapsto\bt_\nu-\bm{s}_\nu
\]
in the  bilinear equation \eqref{bletau}.
Accordingly, the left hand side of the bilinear equation reads
\begin{align}
\hbox{l.h.s.}=&\res_z\left( z^{-1}\tau_1\left(\bt_1+\bm{s}_1-[z^{-1}],\bt_2+\bm{s}_2\right) \tau_2\left(\bt_1-\bm{s}_1+[z^{-1}],\bt_2-\bm{s}_2\right) e^{\xi(2\bm{s}_1;z)} \right)
\nn\\
=& \res_z\left( z^{-1}e^{\xi(2\bm{s}_1;z)} e^{-\sum_{r\ge1}\frac{1}{r z^r}D_{1,r}} e^{\sum_{k\ge1}s_{1,k} D_{1,k}}e^{\sum_{l\ge1}s_{2,l} D_{2,l}} \right)\tau_1\cdot\tau_2
\nn\\
=& \sum_{j\ge0} p_j(2\bm{s}_1)p_{j}(-\widetilde{\bs{D}}_{1}) \sum_{\bs{m},\bs{n}\in\mathcal{I} }
\frac{{\bm{s}_1}^{\bs{m}}\,{\bm{s}_2}^{\bs{n}} }{\bs{m}!\,\bs{n}!} {\bs{D}_1}^{\bs m}{\bs{D}_2}^{\bs n} \,\tau_1\cdot\tau_2
\nn\\
=& \sum_{\bs{l}\in\mathcal{I} } \frac{ (2\bm{s}_1)^{\bs l} }{ \bs{l}! } p_{||\bs l||}(-\widetilde{\bs{D}}_{1}) \sum_{\bs{m},\bs{n}\in\mathcal{I} }
\frac{{\bm{s}_1}^{\bs{m}}\,{\bm{s}_2}^{\bs{n}} }{\bs{m}!\,\bs{n}!} {\bs{D}_1}^{\bs m}{\bs{D}_2}^{\bs n} \,\tau_1\cdot\tau_2
\nn\\
=& \sum_{\bs{m},\bs{n},\bs{l}\in\mathcal{I} } \frac{ 2^{|\bs l|} }
{ \bs{l}!\,\bs{m}!\,\bs{n}! }
{\bm{s}_1}^{\bs{l}+\bs{m}}\,{\bm{s}_2}^{\bs{n}} p_{||\bs l||}(-\widetilde{\bs{D}}_{1})  {\bs{D}_1}^{\bs m}{\bs{D}_2}^{\bs n} \,\tau_1\cdot\tau_2
\nn\\
=&  \sum_{\bs{m},\bs{n}\in\mathcal{I} } \sum_{\bs{l}\le\bs{m}} \frac{2^{|\bs{l}|}}{(\bs{m}-\bs{l})!\,\bs{l}!\, \bs{n}!}  {\bm{s}_1}^{\bs{m}}\,{\bm{s}_2}^{\bs{n}} p_{||\bs{l}||}(-\widetilde{\bs{D}}_1){\bs{D}_1}^{\bs{m}-\bs{l}} {\bs{D}_2}^{\bs{n}}\,\tau_1\cdot\tau_2, \label{hirlhs}
\end{align}
where the third equality is due to
\[
\sum_{j\ge0} p_j(2\bm{s}_1)=\sum_{j\ge0}\sum_{||\bs l||=j} \frac{ (2\bm{s}_1)^{\bs l} }{\bs{l}!}=\sum_{\bs{l}\in\mathcal{I} } \frac{ (2\bm{s}_1)^{\bs l} }{ \bs{l}! }.
\]
In the same way, the right hand side of the bilinear equation is
\begin{align}
\hbox{r.h.s.}=&\res_z\left( z^{-1}\tau_1\left(\bt_1-\bm{s}_1,\bt_2-\bm{s}_2+[z^{-1}]\right) \tau_2\left(\bt_1+\bm{s}_1,\bt_2+\bm{s}_2-[z^{-1}]\right) e^{\xi(2\bm{s}_2;z)} \right)
\nn\\
=& \res_z\left( z^{-1}e^{\xi(2\bm{s}_2;z)} e^{\sum_{r\ge1}\frac{1}{r z^r}D_{2,r}} e^{-\sum_{k\ge1}s_{1,k} D_{1,k}}e^{-\sum_{l\ge1}s_{2,l} D_{2,l}} \right)\tau_1\cdot\tau_2
\nn\\
=& \sum_{j\ge0} p_j(2\bm{s}_2)p_{j}(\widetilde{\bs{D}}_{2}) \sum_{\bs{m},\bs{n}\in\mathcal{I} }
\frac{ {(-\bm{s}_1)}^{\bs{m}}\,{(-\bm{s}_2)}^{\bs{n}} }{\bs{m}!\,\bs{n}!} {\bs{D}_1}^{\bs m}{\bs{D}_2}^{\bs n} \,\tau_1\cdot\tau_2
\nn\\
=& \sum_{\bs{l}\in\mathcal{I} } \frac{ (2\bm{s}_2)^{\bs l} }{ \bs{l}! } p_{||\bs l||}(\widetilde{\bs{D}}_{2}) \sum_{\bs{m},\bs{n}\in\mathcal{I} }
\frac{ {(-\bm{s}_1)}^{\bs{m}}\,{(-\bm{s}_2)}^{\bs{n}} }{\bs{m}!\,\bs{n}!} {\bs{D}_1}^{\bs m}{\bs{D}_2}^{\bs n} \,\tau_1\cdot\tau_2
\nn\\
=& \sum_{\bs{m},\bs{n},\bs{l}\in\mathcal{I} } \frac{ (-1)^{|\bs m|+|\bs n|}\, 2^{|\bs l|} }
{ \bs{l}!\,\bs{m}!\,\bs{n}! }
{\bm{s}_1}^{\bs{m}}\,{\bm{s}_2}^{\bs{l}+\bs{n}}  {\bs{D}_1}^{\bs m}{\bs{D}_2}^{\bs n} p_{||\bs l||}(\widetilde{\bs{D}}_{2}) \,\tau_1\cdot\tau_2
\nn\\
=& \sum_{\bs{m},\bs{n}\in\mathcal{I} }  \sum_{\bs{l}\le\bs{n}} \frac{(-1)^{|\bs{m}|+|\bs{n}-\bs{l}|}\,2^{|\bs{l}|}}{\bs{m}!\,(\bs{n}-\bs{l})!\,\bs{l}!} {\bm{s}_1}^{\bs{m}}\,{\bm{s}_2}^{\bs{n}} {\bs{D}_1}^{\bs{m}}{\bs{D}_2}^{\bs{n}-\bs{l}} p_{||\bs{l}||}(\widetilde{\bs{D}}_2)\,\tau_1\cdot\tau_2.   \label{hirrhs}
\end{align}
By comparing the coefficients of ${\bm{s}_1}^{\bs{m}}\,{\bm{s}_2}^{\bs{n}} $ in \eqref{hirlhs} and \eqref{hirrhs}, we conclude the proposition.
\end{prf}

\begin{exa}
One can write some equations in \eqref{hir} explicitly as follows:
\begin{itemize}
\item[(i)] For  $\bs{m}=(2,0,0,\dots)$ and $\bs{n}=(0,0,0,\dots)$,
\begin{equation}\label{hir20}
\left(D_{1,2}+{D_{1,1}}^2\right)\tau_1\cdot\tau_2=0;
\end{equation}
\item[(ii)] For $\bs{m}=(0,0,0,\dots)$ and $\bs{n}=(2,0,0,\dots)$,
\begin{equation}\label{hir02}
\left(D_{2,2}-{D_{2,1}}^2\right)\tau_1\cdot\tau_2=0;
\end{equation}
\item[(iii)] For $\bs{m}=(3,0,0,\dots)$ and $\bs{n}=(0,0,0,\dots)$,
\begin{equation}\label{hir30}
\left(4 D_{1,3}+3 D_{1,2}D_{1,1}-{D_{1,1} }^3 \right) \, \tau_1\cdot\tau_2=0.
\end{equation}
\item[(iv)] For $\bs{m}=(4,0,0,\dots)$ and $\bs{n}=(0,0,0,\dots)$,
\begin{equation}\label{hir40}
\left(6 D_{1,4}+8 D_{1,3}D_{1,1}-3{D_{1,2} }^2 + {D_{1,1} }^4 \right) \, \tau_1\cdot\tau_2=0.
\end{equation}
\item[(v)] For $\bs{m}=(2,1,0,0,\dots)$ and $\bs{n}=(0,0,0,\dots)$,
\begin{equation}\label{hir210}
\left(2 D_{1,4} +{D_{1,2} }^2 -  D_{1,2}{D_{1,1}}^2 \right) \, \tau_1\cdot\tau_2=0.
\end{equation}
\item[(vi)] For $\bs{m}=(0,0,1,0,0,\dots)$ and $\bs{n}=(0,2,0,0,\dots)$,
\begin{align*}\label{}
&\left( 6 D_{1,3} D_{2,4}+ 8 D_{1,3} D_{2,1} D_{2,3}- D_{1,3} {D_{2,2}}^2+
 6 D_{1,1} D_{1,2} {D_{2,2}}^2\right. \nn\\
&\qquad \left. - 2 D_{1,1}^3 {D_{2,2}}^2 - 6 D_{1,3} {D_{2,1}}^2 D_{2,2} +  D_{1,3} {D_{2,1}}^4
\right) \, \tau_1\cdot\tau_2=0.
\end{align*}
\end{itemize}
Observe that equations \eqref{hir20}--\eqref{hir210} agree with those of the mKP hierarchy given in \cite{JM1983}.
In particular, with
\[
\beta=\log\frac{\tau_2}{\tau_1}, \quad u=\frac{\p^2\log\tau_1}{\p x^2},
\]
 equations \eqref{hir20} and \eqref{hir02} can be recast to equations \eqref{betat12} and \eqref{betat22} respectively. Moreover, if one rewrites $\tau_1=\tau_{KW}$ and $\tau_2=\tau_{o}$, then equations \eqref{hir20} and \eqref{hir30}--\eqref{hir210} coincide with the Hirota equations given in Section~3 of \cite{Ale} to describe the intersection numbers on the moduli space of Riemann surfaces with boundary (see also \cite{Bur2015,Bur2016}).
\end{exa}

\section{Concluding remarks}

With the help of pseudo-differential operators of two derivations, we have constructed the KP-mKP hierarchy \eqref{exPhinut1}--\eqref{exPhinumu}, which can be reduced to the KP hierarchy, the mKP hierarchy and the two-component BKP hierarchy. From our approach, as well as those in \cite{GHW2023,Shi}, one sees that pseudo-differential operators of more derivations than one provide an efficient tool to describe integrable hierarchies.

The KP-mKP hierarchy \eqref{exPhinut1}--\eqref{exPhinumu} is rewritten into two versions of bilinear equations, say, equations \eqref{exble} and \eqref{exble3}, of Baker-Akhiezer functions and adjoint Baker-Akhiezer functions. Based on the former version of these bilinear equations, we show that the KP-mKP hierarchy is equivalent to a certain subhierarchy of the dispersive Whitham hierarchy. Based on the latter version of these bilinear equations, we obtain Hirota equations of two tau functions of the KP-mKP hierarchy, part of which are satisfied by generating functions of intersection numbers. We hope that our results would be helpful to understand the dispersive Whitham hierarchy of general cases, and generating functions for the intersection numbers on the moduli space of Riemann surfaces with boundary. We will study these topics elsewhere.

{\bf Acknowledgments.}
{\noindent \small The first author thanks Jipeng Cheng for useful discussions. This work is partially supported by NSFC No.\,12022119 and No.\,11831017. }


\end{document}